\documentclass[a4paper,11pt]{article}
\usepackage{jheppub}
\usepackage{duckuments}
\usepackage{jheppub}
\usepackage[T1]{fontenc}
\usepackage[utf8]{inputenc}
\usepackage{lipsum}
\usepackage{amsmath,amssymb,amsfonts,bbold}
\usepackage{braket}
\usepackage{natbib}
\usepackage[dvipsnames,svgnames,x11names]{xcolor}
\usepackage{graphicx,subfigure}
\usepackage{hyperref}
\usepackage{mathtools}
\usepackage[normalem]{ulem}
\newcommand{\be}{\begin{equation}}
\newcommand{\ee}{\end{equation}}
\newcommand{\bse}{\begin{subequations}}
\newcommand{\ese}{\end{subequations}}
\newcommand{\ba}{\begin{eqnarray}}
\newcommand{\ea}{\end{eqnarray}}
\newcommand{\bea}{\begin{eqnarray}}
\newcommand{\eea}{\end{eqnarray}}

\newcommand{\lb}{\left (}
\newcommand{\rb}{\right )}
\newcommand{\mL}{\mathcal{L}}
\begin{document}

%\preprint{}

\title{Decoding multiway gravitational junctions in AdS in terms of holographic quantum maps}
\author[1]{Avik Chakraborty,}
\emailAdd{avik.phys88@gmail.com}
\affiliation[1]{Departamento de Ciencias F\'isicas, Facultad de Ciencias Exactas, Universidad Andres Bello,
Sazi\'e 2212, Piso 7, Santiago, Chile}
\author[2,3]{Tanay Kibe,}
\emailAdd{tanay.kibe@ib.edu.ar }
\affiliation[2]{Instituto Balseiro, Centro 
At{\'o}mico Bariloche, S.C. de Bariloche, 8400, 
R{\'i}o Negro, Argentina}
\affiliation[3]{National Institute for Theoretical and Computational Sciences,
School of Physics and Mandelstam Institute for Theoretical Physics,
University of the Witwatersrand, Wits, 2050, South Africa}
\author[4,5]{Mart\'in Molina,}
\emailAdd{martinmolinaramos95@gmail.com}
\affiliation[4]{Departamento de F\'{\i}sica, Universidad T\'{e}cnica Federico Santa Mar\'{\i}a,
Casilla 110-V, Valpara\'{\i}so, Chile,}
\author[5]{Ayan Mukhopadhyay,}
\emailAdd{ayan.mukhopadhyay@pucv.cl}
\affiliation[5]{Instituto de F\'{\i}sica, Pontificia Universidad Cat\'{o}lica de Valpara\'{\i}so,
Avenida Universidad 330, Valpara\'{\i}so, Chile}
\author[6]{and Giuseppe Policastro}
\emailAdd{giuseppe.policastro@phys.ens.fr}
\affiliation[6]{Laboratoire de Physique de l'\'{E}cole Normale Supérieure, ENS, Universit\'{e} PSL, CNRS, Sorbonne Universit\'{e}, Universit\'{e} de Paris, F-75005 Paris, France}

%\date{\today}

\abstract{It has been shown that multiway junctions gluing $n$ copies of locally AdS$_3$ spacetimes ($n\geq 2$) can be described by $n-1$ strings obeying non-linear Nambu-Goto equations coupled by Monge-Amp\`{e}re like terms. Here we study how such junctions along with their stringy degrees of freedom can be interpreted in terms of an interface between $n$ identical holographic conformal theories each defined on a semi-infinite line (wire). We study the gravitational scattering problem at the multiway junction, and show that at the linearized order the dual interfaces correspond to quantum maps which factorize into a product of a scattering matrix determined only by the tension of the dual junction and relative automorphisms of the Virasoro algebra governed by the $n-1$ stringy modes. Both of these are universal in the sense that they are independent of linear modifications of the background state. These generalize earlier results for the 2-way junctions implying that the dual interface is a tunable energy transmitter. We comment on understanding the quantum map corresponding to the full non-linear gravitational problem, and study Ward identities and unitarity bounds.}

\maketitle

\section{Introduction} 
The holographic duality \cite{Maldacena:1997re,Gubser:1998bc,Witten:1998qj} enables the reformulation of quantum gravity in terms of a non-gravitating quantum field theory living at the boundary of spacetime. 
Holographic bulk reconstruction has been fundamental for understanding how spacetime is encoded in the dual field theory. Concretely, this program has shown that bulk locality and the emergence of spacetime should be thought of in terms of a quantum error correcting code that encodes bulk operators into the boundary theory (see \cite{harlow2018tasi,Jahn:2021uqr,Chen:2021lnq,Kibe:2021gtw} for reviews).
Ultraviolet complete examples of holography exist only within the framework of string theory, where both matter and spacetime arise as quantized vibrations of a fundamental string \cite{Green:2012oqa,Green:1987mn,Polchinski:1998rq,Polchinski:1998rr}. Moreover, dynamical extended objects in the gravitational theory, such as strings and branes, are essential for its non-perturbative completion. Therefore, understanding how such extended objects are encoded in the structure of the dual field theory is of fundamental importance. The first steps towards this goal can be taken in the large $N$ and strong coupling limit of the field theory, which corresponds to a classical gravitational theory coupled to a few fields.

Recently, it was shown that the Nambu-Goto equation arises from gravitational junction conditions for a junction formed by gluing two three-dimensional anti-de Sitter (AdS$_3$) spacetimes \cite{Banerjee:2024sqq}. Such gravitational junctions are holographic models for interfaces in conformal field theory (CFT) \cite{Karch:2000ct,Karch:2001cw,DeWolfe:2001pq,Bachas:2001vj}.Two-way gravitational junctions correspond to interfaces that are formed by joining two wires at a point, each of which is a strongly coupled $1+1$ dimensional CFT with a large central charge and sparse spectrum. The tension of the junction (string) characterizes the defect operator at the dual interface. 

There has been some recent progress in interpreting the stringy Nambu-Goto excitations of the gravitational junction in the dual conformal interface between two identical CFTs, each in the same background state. It has been shown in \cite{Chakraborty:2025dmc} that each stringy excitation of the junction between two locally $AdS_3$ spacetimes corresponds to a quantum map $\mathcal{H}_{\rm in} \to \mathcal{H}_{\rm out}$ from the Hilbert space {(in the universal sector)} of the incoming excitations to that of the outgoing excitations at the conformal interface. Concretely, this map is a composition of a universal scattering with a one-sided automorphism of the Virasoro algebra that is parametrized by the stringy modes and which redistributes energy in the in or out Hilbert space. The one-sided automorphism, at the linear order in the incoming/outgoing energy modes, arises simply due to a one-sided conformal transformation of the wires. Equivalently, the stringy excitations can be translated in terms of quantum maps $\mathcal{H}_1\to \mathcal{H}_2$ from the Hilbert space (in the universal sector) of one CFT to the other. Due to the presence of the stringy modes, the usual defect operator is similarly generalized by a one-sided automorphism of the Virasoro algebras. Furthermore, these quantum maps were shown to be independent of the choice of background state of the CFTs, again to linear order in the energy modes.

The above developments were in the context of two-way conformal interfaces. In this paper, we are concerned with understanding how the quantum maps generalize to the setting of interfaces formed by joining multiple conformal wires, each in the same background state, at a point. These interfaces correspond to a gravitational junction gluing $n\geq2$ locally AdS$_3$ spacetimes. It has been shown that the general solutions of the $n$-way gravitational junctions correspond to $n-1$ strings that obey non-linear Nambu-Goto equations coupled by Monge-Amp\'{e}re like terms \cite{Chakraborty:2025jtj}. For $n\geq 3$, non-trivial solutions to the gravitational junction equations persist even in the tensionless limit. Remarkably, this is a demonstration of how matter-like behavior emerges out of pure gravity. 

Here we study gravitational scattering at the multi-way junction, and we find that at the linearized order the stringy excitations of the multi-way gravitational junction can be translated to quantum maps in the dual conformal interface. The $n$-way interface acts as a quantum map $\mathcal{H}_{\rm in} \to \mathcal{H}_{\rm out}$, from the space of incoming to that of the outgoing excitations at the interface, generalizing the result of \cite{Chakraborty:2025dmc} for $n=2$. This $n \rightarrow n$ map is the composition of a universal scattering with conformal transformations of $n-1$ wires, which are parametrized by the stringy excitations of the gravitational junction. The full map is also independent of the choice of the background state. The conformal transformations result in an energy re-distribution in the in or out space. 

Our results indicate that the interface is a tunable energy transmitter. For any set of values of the incoming energy modes, we can find appropriate stringy modes for which the interface is purely reflexive {(factorized)}. Similarly, one can find a pseudo-topological limit for any set of values of the incoming energy modes as discussed below. This generalizes the results for the two-way case discussed in \cite{Chakraborty:2025dmc}.

We also show that the usual Ward identities for a two-way conformal interface \cite{Chakraborty:2025dmc} are generalized to the multi-way setting. The Ward identities are best viewed via bi-partitions of the wires such that one of the partitions contain only a single wire \cite{Chakraborty:2025jtj}, say the $i^{\rm th}$ one. The sources of the energy-conservation Ward identities vanish at the interface due to the conformal boundary condition. Furthermore, the sources for the momentum-conservation Ward identities are determined by the stringy excitations of the dual junction. The $i^{\rm th}$ source corresponds to the expectation value of a generalized displacement operator that measures the energy cost of displacing the $i^{\rm th}$ wire away from the interface. The stringy modes can be chosen so that the expectation value of the {maximal number} of generalized displacement operators vanishes, which corresponds to a pseudo-topological limit in which the { energy scattering matrix becomes an involution.}

The rest of this paper has been structured as follows. In Sec.~\ref{Sec:jn_soln}, we construct the general solutions of the gravitational junction between $n$ locally AdS$_3$ spacetimes in the presence of incoming and outgoing gravitational excitations by perturbing about the exact permutation symmetric static solution. We then interpret the $n$-way junctions in terms of an $n$-way conformal interface. This is followed by Sec.~\ref{Sec:holqmap}, where we show that the stringy modes of the gravitational junction can be interpreted in terms of quantum maps corresponding to the dual multi-way conformal interface. We also demonstrate that these maps are tunable and present a generalization of the Ward identities. Furthermore, we discuss the interpretation of the junction as a boundary state in the interface CFT which can be expected to hold even at higher orders in the perturbative expansion. We conclude with a summary and discussion of future directions in Sec.~\ref{Sec:conclusions}. Appendix \ref{Sec:assymetricsoln} describes permutation asymmetric exact static solutions of the gravitational junction. We study the perturbations and show that the corresponding quantum maps violate unitarity and thus we justify why we have discarded these asymmetric solutions in our analysis.

\section{Gravitational multiway junctions and linearized scattering} \label{Sec:jn_soln}

\subsection{Review of multiway junctions gluing copies of identical spacetimes}
\begin{figure}
    \subfigure[]{\includegraphics[width=0.5\textwidth]{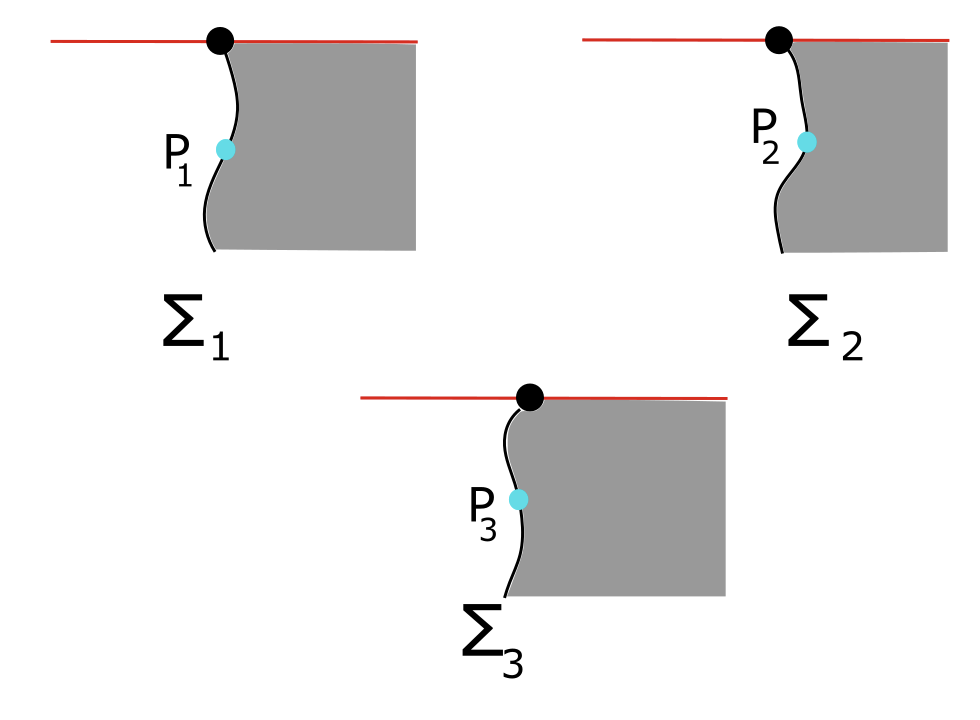}}
    \subfigure[]{\includegraphics[width=0.5\textwidth]{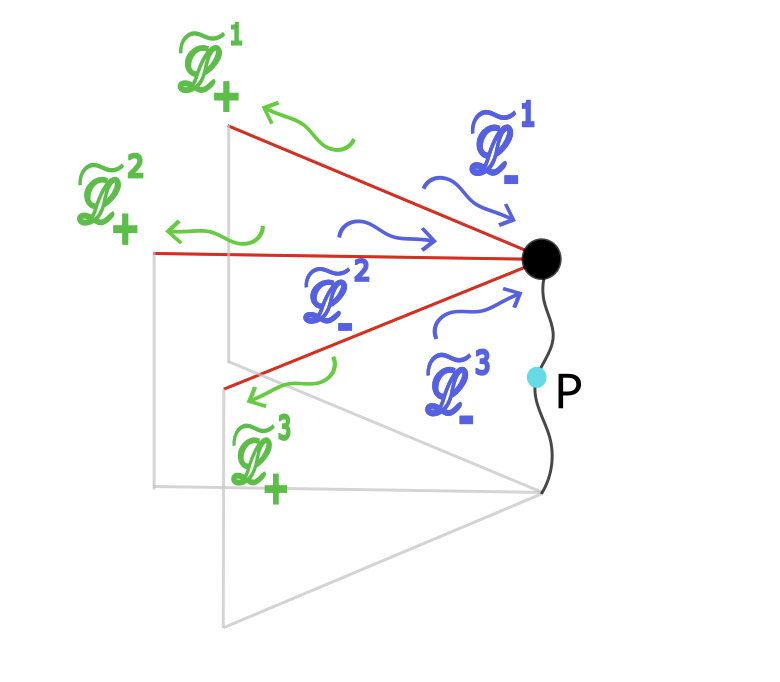}}
    \caption{A three-way junction: (a) Three asymptotically AdS$_3$ spacetimes with the gluing hypersurfaces $\Sigma_{1,2,3}$. The gray region is excised. (b) The gravitational junction with the incoming (blue) and outgoing energy fluxes shown. The red lines are the holographic CFTs and the black dot is the conformal interface. We identify the points $P_i$ (in cyan) on $\Sigma_i$ with the point $P$ on $\Sigma$ in (b). Incoming (blue) and outgoing (green) energy fluxes are shown on each of the CFTs.}
    \label{Fig:setup}
\end{figure}

In this section, we follow the exposition of multi-way junction conditions in \cite{Chakraborty:2025jtj}. {For illustration, a three-way junction is shown in Fig.~\ref{Fig:setup}.} Consider $n$ \textit{identical} copies $\mathcal{M}_i$ (with $i=1,\cdots,n$) of a locally AdS$_3$ manifold $\mathcal{M}$, each of which is divided into two halves, $\mathcal{M}_{iL}$ and $\mathcal{M}_{iR}$, by distinct co-dimension one hypersurfaces $\Sigma_i$. A gravitational junction $\Sigma$ is formed by gluing $n$ such fragments $\mathcal{M}_{i\alpha_i}$, with $\alpha_i=L,R$, resulting in the full spacetime $\widetilde{\mathcal{M}}$ that satisfies Einstein's equations along with the required junction conditions. Each point $P$ in the junction $\Sigma$ is constructed by identifying the corresponding points $P_i$ in $\Sigma_i$. Hence, each $\Sigma_i$ should be considered as the image of $\Sigma$ in the corresponding $\mathcal{M}_i$. These \textit{identifications} of the points $P_i$ of $\Sigma_i$ and the \textit{embeddings} of $\Sigma_i$ in $\mathcal{M}_i$ should satisfy the gravitational junction conditions \cite{Israel:1966rt} at $\Sigma$. Let $t$ be the time and $z,x$ be the spatial coordinates of $\mathcal{M}$. Since all the $n$ copies $\mathcal{M}_i$ inherit the coordinate charts of $\mathcal{M}$, the fragments $\mathcal{M}_{\alpha_i}$ have coordinates $(t_i,z_i,x_i)$. The embeddings of $\Sigma_i$ in $\mathcal{M}_i$ are specified by the $n$ functions
\begin{equation}\label{Eq:embed}
  \Sigma_i:\,\,  x_i = f_{i}(t_i, z_i)\,, \ \ i = 1, 2,\cdots,n\,,
\end{equation}
where $x_i$ are the coordinates transverse to $\Sigma_i$. We form the junction $\Sigma$ by identifying the points $P_i$ to a point $P$ in $\Sigma$, to which we assign the worldsheet coordinates $(\tau,\sigma)$. We fix the worldsheet gauge by choosing the coordinates $(\tau,\sigma)$ which satisfy
\begin{align}\label{Eq:WSgauge}
    \tau(P) = \frac{1}{n}\sum_{i=1}^n t_i(P_i), \quad \sigma(P) = \frac{1}{n}\sum_{i=1}^n z_i(P_i)\,.
\end{align}
Then we are left with $2(n-1)$ independent variables, $t_i - t_j$ and $z_i - z_j$ ($i\neq j$), which are the relative shifts of time and space, respectively, as we move from $\mathcal{M}_{i\alpha_i}$ to $\mathcal{M}_{j\beta_j}$ across the junction $\Sigma$. Therefore, together with the $n$ embedding functions $f_i$ of $\Sigma_i$, we have in total $3n-2$ variables that completely specify the junction. Note that all variables are functions of the worldsheet coordinates $\tau$ and $\sigma$.

The full gravitational action, which determines the bulk metric and gives the junction conditions, is 
\begin{equation}\label{Eq:bulk-action}
    S=\frac{1}{16\pi G_N}\int_{\mathcal{\widetilde{\mathcal{M}}}}d^3x \sqrt{-g}(R - 2\Lambda)+T_0\int_{\Sigma}{\rm d}\tau{\rm d}\sigma\,\sqrt{-\gamma}+{\rm GHY \ terms}\,,
\end{equation}
where $g$ is the \textit{only} degree of freedom, GHY are the Gibbons-Hawking-York boundary terms and $T_0$ is the tension of the string constituting the junction. {Note that $\Sigma_i$ have distinct GHY terms as they have different embeddings, and other boundaries do not contribute to the junction conditions}. Varying the action away from the junction $\Sigma$ implies 
\begin{equation} \label{Eq:Ein}
    R_{MN}-\frac{1}{2}Rg_{MN}+\Lambda g_{MN}=0\,,
\end{equation} 
that is, each fragment $\mathcal{M}_{i \alpha_i}$ is an Einstein manifold. The first junction condition, which has been assumed in the gravitational action \eqref{Eq:bulk-action}, states that the induced metric is continuous and therefore the worldsheet metric $\gamma$ is
\begin{align}\label{Eq:hcont}
&\gamma_{\mu\nu}(\tau, \sigma) := \gamma_{1,\mu\nu}(\tau,\sigma)=\cdots = \gamma_{n,\mu\nu}(\tau,\sigma)\,.
\end{align}
The variation of the action \eqref{Eq:bulk-action} with respect to $g$ at the junction gives
\begin{equation}\label{Eq:Kdisc}
    \sum_{i=1}^n (-1)^{s(\alpha_i)}\left(K_{i,\mu\nu} - K_i\,\gamma_{i,\mu\nu}\right) = 8\pi G_N T_0 \gamma_{\mu\nu}\,,
\end{equation}
with $s(\alpha_i) = 0$ if $\alpha_i = L$ and $s(\alpha_i) = 1$ if $\alpha_i = R$. Here, $K_{i,\mu\nu}$ is the extrinsic curvature of $\Sigma_i$ in $\mathcal{M}_{i\alpha_i}$ and $K_i = \gamma^{\mu\nu}K_{i,\mu\nu}$. The bulk diffeomorphism symmetry implies that the \textit{total} Brown-York tensor of the junction, which is the left hand side of \eqref{Eq:Kdisc}, is conserved. Therefore, we obtain only one independent equation from \eqref{Eq:Kdisc}, which together with $3(n-1)$ equations from \eqref{Eq:hcont} give $3n-2$ independent equations, exactly matching the number of unknown variables. For simplicity, we assume that $\alpha_i =L$ for all $i$. Then, following \cite{Chakraborty:2025jtj}, we define $3n-3$ independent relative shifts of the time ($\tau_{d_i}$), the radial coordinate ($\sigma_{d_i}$), and the transverse coordinate ($x_{d_i}$), across the junction as
\begin{eqnarray}\label{Eq:xsxd}
\tau_{d_i} &=&\begin{cases}
        \frac{1}{n}(t_n-t_{i+1}) \,\, {\rm for}\,\, i = 1, \cdots, n-2\\
        \frac{1}{n}(t_n-t_1) \,\, {\rm for}\,\, i = n-1
    \end{cases}\,,\nonumber\\
\sigma_{d_i} &=&\begin{cases}
        \frac{1}{n}(z_n-z_{i+1}) \,\, {\rm for}\,\, i = 1, \cdots, n-2\\
        \frac{1}{n}(z_n-z_1) \,\, {\rm for}\,\, i = n-1
    \end{cases}\,,\nonumber\\
x_{d_i} &=&\begin{cases}
        \frac{1}{n}(x_n-x_{i+1}) \,\, {\rm for}\,\, i = 1, \cdots, n-2\\
        \frac{1}{n}(x_n-x_1) \,\, {\rm for}\,\, i = n-1
    \end{cases}\,.
\end{eqnarray}
Equation \eqref{Eq:xsxd} along with the averaged transverse coordinate
\begin{equation}
    x_s = \frac{1}{n}\sum_i x_{i}.
\end{equation} 
give the necessary $3n-2$ functions of $\tau$ and $\sigma$ that we need to determine. If a subset of the  $n$ fragments, $\mathcal{M}_{i\alpha_i}$ are $\mathcal{M}_{iR}$ instead of $\mathcal{M}_{iL}$, we simply reverse the sign of the transverse coordinate $x_i$ in the parameterization \eqref{Eq:xsxd} for the values of $i$ in this subset. We will look for solutions that satisfy the Dirichlet boundary conditions
\begin{equation}\label{Eq:DBC}
    \lim_{\sigma\rightarrow 0}x_i = 0\Rightarrow \lim_{\sigma\rightarrow 0}x_s = 0, \,\,\lim_{\sigma\rightarrow 0}x_{d_i} =0
\end{equation}
at the boundary of AdS. Solutions with the above boundary conditions can be interpreted as an $n-$way interface in the dual CFT, i.e $n$ conformal wires joined at $x=0$ (see Sec. \ref{Sec:holoint}).

It has been shown in \cite{Banerjee:2024sqq,Chakraborty:2025jtj} that generic solutions of the $n-$way junction correspond to coupled $n-1$ strings, upto $3n$ rigid parameters related to spacetime and worldsheet isometries. 

{Particularly, any solution of the multiway junction conditions gluing $n$ \textit{identical} copies of $\mathcal{M}$  has the following properties.}
\begin{enumerate}
    \item The $n-1$ hypersurfaces 
        \begin{equation*}
    \Sigma_{NG_i}: t=\tau, \,\, z=\sigma,\,\, x = x_{d_i}(\tau,\sigma)
        \end{equation*}
        correspond to solutions to the {\textit{non-linear}} Nambu-Goto equations for their embeddings in $\mathcal{M}$ coupled by ({\textit{non-linear}}) Monge-Amp\`{e}re like terms. 
    \item $x_{d_i}$ are the only degrees of freedom, implying that $x_s$, $\tau_{d_i}$ and $\sigma_{d_i}$ are completely determined as functions of $\tau$, $\sigma$ and the tension for any given choice of the solution of the coupled Nambu-Goto equations.
    \item {For $n=2$, the full spacetime obtained as a result of the gluing at the junction is smooth (a manifold) when \textit{both} the tension $T_0$ and the rigid parameters vanish.}
    \item {For $n\geq 3$, the degrees of freedom described by the coupled Nambu-Goto equations survive in the limit in which \textit{both} the tension $T_0$ and the rigid parameters vanish. Note that the full spacetime obtained as a result of the gluing is never smooth (a manifold) in this case.}
\end{enumerate}
Remarkably, the last feature in the above list implies that multiway junctions gluing three dimensional spacetimes provide a setup in which matter like vibrations can arise from \textit{pure} gravity. Generalizations to higher dimensional setups have been discussed in \cite{Chakraborty:2025jtj}.

\subsection{Linearized scattering at multiway junctions}
{Here we generalize results of \cite{Chakraborty:2025jtj} to junctions gluing \textit{non-identical} locally AdS$_3$ spacetimes with the aim of studying linearized scattering.} We consider a gravitational junction between $n\geq 3$ fragments of  locally AdS$_3$ manifolds, each of which is a {\textit{distinct}} Ba\~nados spacetime \cite{Banados:1998gg} ($\mathcal{M}$) endowed with the metric 
\begin{multline}\label{Eq:Banados-Metric}
    ds^2= \frac{dz^2}{z^2}+2dtdx\left(\mL_+(x^+)-\mL_-(x^-)\right)\\- \frac{dt^2}{z^2}\left(1-z^2\mL_+(x^+)\right)\left(1-z^2\mL_-(x^-)\right)\\+\frac{dx^2}{z^2}\left(1+z^2\mL_+(x^+)\right)\left(1+z^2\mL_-(x^-)\right),
\end{multline}
where $x^\pm=t\pm x$. For simplicity, we have set the cosmological constant $\Lambda = -1$ in all the spacetimes glued at the junction. For future purposes, it is also useful to define the dimensionless tension $\lambda=8\pi G_N T_0$. 

{Our goal, which is going to be realized in the following section, is to decode the solutions of the gravitational junction conditions in terms of quantum maps at the dual multi-interface generalizing the results of the two-way junction studied in \cite{Bachas:2020yxv,Chakraborty:2025dmc,Banerjee:2025}. For this purpose, it is useful to obtain the dependence of the quantum maps on $\lambda$ exactly as in \cite{Bachas:2020yxv,Chakraborty:2025dmc,Banerjee:2025} and so we do not proceed as in \cite{Banerjee:2024sqq,Chakraborty:2025jtj} where $\lambda$ was treated to be a small parameter. Nevertheless, with the aim of studying linearized scattering we will assume that the departure from the exact static solution(s) of the multiway junction conditions to be small as in \cite{Chakraborty:2025dmc,Banerjee:2025}. Thus we will assume that the departures of the Ba\~nados spacetimes glued at the junction from Poincar\'{e} patch AdS$_3$ are small, and also the amplitudes of the vibrations of Nambu-Goto modes constituting the degrees of freedom of the junction to be small. }

Accordingly, we proceed by assuming that
\begin{equation}\label{Eq:Lj}
    \mathcal{L}_{\pm}^{(j)}(x_i^\pm)= \mathcal{L}^{j}_{\omega,\pm} e^{i \omega x_{i}^\pm}\,\,\, {\rm with}\, \,\,\mathcal{L}^{j}_{\omega,\pm} =\mathcal{O}(\epsilon)
\end{equation}
for $j=1,...,n$ in the metrics \eqref{Eq:Banados-Metric} of $\mathcal{M}_{j}$. We will solve the junction conditions \eqref{Eq:hcont} and \eqref{Eq:Kdisc} constituting $3n-2$ equations for $3n-2$ variables (as discussed in the previous section) first exactly at $\mathcal{O}(\epsilon^0)$ to obtain a static solution and then \textit{generally} at $\mathcal{O}(\epsilon)$. Since our analysis will be linear in the latter case, we can assume plane wave forms of $\mathcal{L}_{\pm}^{(j)}(x_i^\pm)$ as in \eqref{Eq:Lj} as generally we can superpose these to form wavepackets. 

At $\mathcal{O}(\epsilon^0)$ we get a unique exact permutation-symmetric and static solution of the non-linear junction conditions \eqref{Eq:hcont} and \eqref{Eq:Kdisc} given by
\begin{equation}\label{Eq:order0soln}
    \tau_{d_i} =0\,, \quad \sigma_{d_i}=0\,, \quad x_{d_i}=0\,, \quad x_s=\sigma p_{\lambda,n}\,\,\, {\rm with}\,\,\, p_{\lambda,n}=\frac{\lambda}{\sqrt{n^2-\lambda^2}},
\end{equation}
which is well defined when $0 \leq \lambda < n$. The positivity of $\lambda$ ($T_0$) follows simply from the bulk null energy condition for the stress tensor of the junction. If $\lambda > n$, the induced metric on the junction is dS$_2$ instead of AdS$_2$. As shown in the next section, the quantum map in the dual interface obeys unitarity bounds when $0 \leq \lambda < n$ for this permutation-symmetric solution. We also have other exact non-permutation symmetric static solutions at $\mathcal{O}(\epsilon^0)$ for $0\leq \lambda <n-2$, which, as described in Appendix~\ref{Sec:assymetricsoln}, give rise to non-unitary quantum maps at the dual multi-interface. So, we will discard these permutation-asymmetric solutions although we do not find any explicit bulk argument to discard these solutions.\footnote{When the energy-momentum tensor is localized on hypersurfaces, it is not easy to find a simple bulk energy condition which corresponds to unitarity and other requirements for the dual quantum field theory. For instance, it has been shown in \cite{Kibe:2021qjy,Banerjee:2022dgv,Kibe:2024icu} that the quantum null energy condition can be violated in quenches in holographic theories although the matter localized on the null hypersurfaces in the dual spacetimes satisfy the classical null energy condition.} Note that the case of $\lambda =0$ is beyond the scope of the linearized analysis presented here.

At $\mathcal{O}(\epsilon)$, we obtain the following equations for $x_{d_i}$: 
 \begin{multline} \label{Eq:NG}
     2\sigma p^2_{\lambda,n}n^3\ddot{x}_{d_i} + 2n\lambda^2 \lb 2 x^\prime_{d_i} - \sigma x^{\prime\prime}_{d_i}\rb = \\ i\omega\lambda^2\sigma^3 e^{i\omega \tau} \lb (\mathcal{L}_{\omega,+}^{n-i}-\mathcal{L}_{\omega,+}^{n})e^{ip_{\lambda,n}\omega\sigma} - (\mathcal{L}_{\omega,-}^{n-i}-\mathcal{L}_{\omega,-}^{n})e^{-ip_{\lambda,n}\omega\sigma} \rb,
 \end{multline}
 where dot and prime denote $\partial_{\tau}$ and $\partial_{\sigma}$ respectively. The left hand side of Eq.\eqref{Eq:NG} is just the linearized Nambu-Goto (NG) equation in empty AdS$_3$ when $\lambda\rightarrow 0$. The sources of this modified linearized NG equation are proportional to $(\mathcal{L}_{\omega,+}^{n-i}-\mathcal{L}_{\omega,+}^{n})$ and $(\mathcal{L}_{\omega,-}^{n-i}-\mathcal{L}_{\omega,-}^{n})$. Note that these sources always vanish when $\omega\rightarrow 0$. Since the amplitudes are treated as small parameters, the non-linear Monge-Amp\`{e}re like terms do not appear.\footnote{The latter is a limitation as we cannot explicitly analyze the role of the degrees of freedom in the tensionless limit. These arise solely from the non-linear Monge-Amp\`{e}re like terms. We say more about this in the concluding section.}

The solutions of the 3n-3 variables $x_{d_i}$, $\tau_{d_i}$ and $\sigma_{d_i}$ are obtained at $\mathcal{O}(\epsilon)$ from perturbative expansion of the metric continuity conditions given by \eqref{Eq:hcont}. We have the following general solution to \eqref{Eq:NG}
 \begin{align} \label{Eq:NG_sol}
    x_{d_i}&=\frac{e^{i\omega\tau}}{2n\omega^3}\Bigg[\sqrt{\frac{1}{\pi}}\left(-\frac{2\lambda\omega}{n^{\frac{1}{3}}p_{\lambda,n}}\right)^\frac{3}{2}
    \nonumber \\ &\quad \times\Bigg(\left(\mathcal{A}_{\omega,1}^i+\frac{np_{\lambda,n}\omega\sigma}{\lambda}\mathcal{A}_{\omega,2}^i\right)\sin \left(\frac{np_{\lambda,n}\omega\sigma}{\lambda}\right) + \left(\mathcal{A}_{\omega,2}^i-\frac{np_{\lambda,n}\omega\sigma}{\lambda}\mathcal{A}_{\omega,1}^i\right)\cos \left(\frac{np_{\lambda,n}\omega\sigma}{\lambda}\right)\Bigg) \nonumber \\
    &\quad \qquad + \left(\mathcal{L}_{\omega,+}^{n}-\mathcal{L}_{\omega,+}^{n-i}\right)e^{ip_{\lambda,n}\omega\sigma}\left(2p_{\lambda,n}\omega\sigma + i(2+\omega^2 \sigma^2)\right)\nonumber \\ &\qquad \qquad+ \left(\mathcal{L}_{\omega,-}^{n}-\mathcal{L}_{\omega,-}^{n-i}\right)e^{-ip_{\lambda,n}\omega\sigma}\left(2p_{\lambda,n}\omega\sigma - i(2+\omega^2 \sigma^2)\right)\Bigg],
\end{align}
where the second line is the general homogeneous solution of the (source-free) linearized Nambu-Goto equation in AdS$_3$. Imposing ingoing boundary conditions \cite{Son:2002sd,Herzog:2002pc,Skenderis:2008dg} at the Poincar\'{e} horizon on the worldsheet, we obtain that
\begin{equation} \label{Eq:ingoing_bc}
\mathcal{A}_{\omega,1}^j=\mathcal{A}_{\omega,nn}^j+\mathcal{A}_{\omega,n}^j, \quad \mathcal{A}_{\omega,2}^j=i \mathcal{A}_{\omega,nn}^j.
\end{equation} 
Above $\mathcal{A}_{\omega,nn}^i$ correspond to non-normalizable modes of the homogeneous NG equation, which are the causal response to bulk perturbations that travel from the boundary towards the Poincar\'{e} horizon of the worldsheet. $\mathcal{A}_{\omega,n}^i$ are intrinsic normalizable (stringy) modes of the homogeneous NG equation.  Both $\mathcal{A}_{\omega,nn}^i$ and $\mathcal{A}_{\omega,n}^i$ are determined by initial and boundary conditions as usual in Lorentzian holographic duality.

The solutions for $\sigma_{d_i}$ and $\tau_{d_i}$ are:
\begin{align}\label{Eq:sdtd-sol}
    \sigma_{d_i}&=\frac{\sigma e^{i\omega\tau}}{n\omega^2}\Bigg[\sqrt{\frac{2}{\pi}}\left(-\frac{\lambda^{\frac{5}{3}}\omega}{np_{\lambda,n}}\right)^{\frac{3}{2}} \Bigg(\mathcal{A}_{\omega,1}^i\cos \left(\frac{np_{\lambda,n}\omega\sigma}{\lambda}\right) -\mathcal{A}_{\omega,2}^i\sin \left(\frac{np_{\lambda,n}\omega\sigma}{\lambda}\right)\Bigg) \notag \\ 
    &\qquad \qquad \qquad \qquad \qquad \qquad \qquad + \left(\mathcal{L}_{\omega,+}^{n-i}-\mathcal{L}_{\omega,+}^{n}\right)e^{ip_{\lambda,n}\omega\sigma} + \left(\mathcal{L}_{\omega,-}^{n-i}-\mathcal{L}_{\omega,-}^{n}\right)e^{-ip_{\lambda,n}\omega\sigma}\Bigg], \\
    \tau_{d_i}&=\frac{ie^{i\omega\tau}}{2n\omega^3}\Bigg[\sqrt{-\frac{1}{\pi}}\left(\frac{2\lambda^{\frac{5}{3}}\omega}{np_{\lambda,n}}\right)^{\frac{3}{2}} \Bigg(\mathcal{A}_{\omega,1}^i\cos \left(\frac{np_{\lambda,n}\omega\sigma}{\lambda}\right) -\mathcal{A}_{\omega,2}^i\sin \left(\frac{np_{\lambda,n}\omega\sigma}{\lambda}\right)\Bigg) \notag \\ 
    &\qquad \qquad \qquad \qquad + \Big(\left(\mathcal{L}_{\omega,+}^{n}-\mathcal{L}_{\omega,+}^{n-i}\right)e^{ip_{\lambda,n}\omega\sigma} + \left(\mathcal{L}_{\omega,-}^{n}-\mathcal{L}_{\omega,-}^{n-i}\right)e^{-ip_{\lambda,n}\omega\sigma}\Big)\left(2-\omega^2\sigma^2\right)\Bigg].
\end{align}
The solution for $x_s$ at $\mathcal{O}(\epsilon)$ which solves the extrinsic curvature discontinuity conditions (recall that only one of these is independent) is given by
\begin{equation} \label{Eq:xssoln}
    x_s=\sigma p_{\lambda,n} + \epsilon x_s^{(1)},
\end{equation}
including the zeroth order part discussed earlier, and where
\begin{align}\label{Eq:xs1}
    x_s^{(1)}=& \frac{e^{i\omega\tau}}{2n\omega^3}\Bigg[e^{ip_{\lambda,n}\omega\sigma}\Big(2p_{\lambda,n}\omega\sigma + i(2+\omega^2\sigma^2)\Big)\left(\sum_{i}\mathcal{L}_{\omega,+}^{i}\right) \nonumber\\&\qquad \qquad+ e^{-ip_{\lambda,n}\omega\sigma}\Big(2p_{\lambda,n}\omega\sigma - i(2+\omega^2\sigma^2)\Big)\left(\sum_{i}\mathcal{L}_{\omega,-}^{i}\right)\Bigg]\,.
\end{align}
In the above solutions we have turned off additional terms that solve the homogeneous equations, since they are not of the plane-wave form and thus are not relevant for the present analysis of linearized scattering.

\subsection{Implementing the Dirichlet boundary condition and preliminary holographic interpretation}\label{Sec:holoint}
The $n-$way gravitational junction is holographically dual to an interface between $n$ identical holographic CFTs living on semi-infinite wires. The CFTs have the light-cone coordinates $x_i^\pm = t_i\pm x_i$, where $t_i$ and $x_i$ are the time and space coordinates, respectively. 
The energy momentum tensors for each of the CFTs can be obtained using holographic renormalization \cite{Henningson:1998gx,Balasubramanian:1999re} as
\begin{equation}
    {T}^{j}_{\pm}(x_j^\pm) = \frac{c\epsilon}{12\pi}e^{i \omega {x}_j^\pm} \mathcal{L}_{\omega,\pm}^{j}\label{Eq:energy1}, \quad j=1,2,\dots ,n.
\end{equation}
Thus, the state on the dual interface CFT corresponds to left and right moving plane-wave excitations on each wire. Note that we have assumed that the gravitational junction is formed by gluing the left halves of $n$ locally AdS$_3$ manifolds. This corresponds to $n$, left half-line CFTs glued at the interface. The right-moving ($\mathcal{L}_{\omega,-}^i$) excitations on each CFT are incoming at the interface and left-moving ($\mathcal{L}_{\omega,+}^i$) excitations are outgoing (see Fig.~\ref{Fig:setup}).

Without loss of generality we assume that the interface is at $x_i=0$. This is imposed on the gravitational solution \eqref{Eq:xssoln} via the Dirichlet boundary conditions
\begin{equation} \label{Eq:Dir_bc}
    \lim_{\sigma\to0} x_s=0, \quad \lim_{\sigma\to0} x_{d_i}=0.
\end{equation}

We readily note from \eqref{Eq:xs1} that the Dirichlet boundary condition $\lim_{\sigma\to0} x_{s}=0$ is satisfied if and only if
 \begin{equation} \label{Eq:energy_con}
     \sum_{i=1}^n \mathcal{L}_{\omega,+}^{i}=\sum_{i=1}^n \mathcal{L}_{\omega,-}^{i},
 \end{equation}
 This is just energy conservation at the interface imposing that the sum of the incoming amplitudes (right hand side) equals the sum of the outgoing amplitudes (left hand side). Equivalently \eqref{Eq:energy_con} is the \textit{conformal boundary condition} of the $n$-way interface as discussed in the next section. Eq. \eqref{Eq:energy_con} can be solved by
 \begin{equation} \label{Eq:param_L_modes}
     \mathcal{L}_{\omega,+}^{i}=\left(1-\sum_{j\neq i}\mathcal{T}_{\omega}^j\right)\mathcal{L}_{\omega,-}^{i} + \mathcal{T}_{\omega}^i\sum_{j \neq i}\mathcal{L}_{\omega,-}^{j}\,, \, i\in\{1,\dots ,n \},
 \end{equation}
 where $\mathcal{T}_{\omega}^i$ and $1-\mathcal{T}_{\omega}^i$ are \textit{arbitrary} coefficients.

We also find from \eqref{Eq:NG_sol} that the Dirichlet boundary conditions $\lim_{\sigma\to0} x_{d_i}=0$ can be used to solve for $\mathcal{A}_{\omega,nn}^i$, in terms of the plane wave amplitudes and the coefficients defined above, as
\begin{align}\label{Eq:Ai_nn_Dir}
     \mathcal{A}_{\omega,nn}^i =& -i\sqrt{\frac{n\pi}{2}}\frac{1}{\left(n^2-\lambda ^2\right)^{3/4} \omega ^{3/2}}\Bigg[\sum_{j=1}^{n-1}\mathcal{L}_{\omega,-}^{j}\Bigg(\mathcal{T}_{\omega}^{n}-\mathcal{T}_{\omega}^{n-i}\Bigg)+\mathcal{L}_{\omega,-}^{n-i}\left(\sum_{j=1}^{n-1}\mathcal{T}_{\omega}^{j}+\mathcal{T}_{\omega}^{n}\right)\nonumber\\
     &\qquad\qquad\qquad\qquad\qquad\qquad\quad-\mathcal{L}_{\omega,-}^{n}\left(\sum_{j=1}^{n-1}\mathcal{T}_{\omega}^{j}+\mathcal{T}_{\omega}^{n-i}\right) \Bigg], \,i=1,\cdots,n-1.
 \end{align}
 After fixing the ingoing and Dirichlet boundary conditions, Eq. \eqref{Eq:ingoing_bc} and Eq. \eqref{Eq:Dir_bc} respectively, $x_{d_i}$ decay as $\sigma^3$ as $\sigma \rightarrow 0$ which has also been observed in \cite{Chakraborty:2025jtj}.

\section{The n-way junction as a holographic quantum map} \label{Sec:holqmap}

An interface between two CFTs can be defined as a linear map from the Hilbert space $\mathcal{H}_1$ of the first CFT to the Hilbert space $\mathcal{H}_2$ of the second CFT. For generalizing to an interface between $n$-CFTs it is more useful to see the two-way interface as a linear map from $\mathcal{H}_{\rm in}$ to $\mathcal{H}_{\rm out}$, where $\mathcal{H}_{\rm in}$ is the tensor product of right-moving excitations of the left CFT and the left-moving excitations of the right CFT constituting the incoming excitations at the interface, and $\mathcal{H}_{\rm out}$ is the tensor product of left-moving excitations of the left CFT and the right-moving excitations of the right CFT constituting the outgoing excitations at the interface. In the general $n$-way interface, it will be convenient to consider the \textit{folded} picture in which we think of the interface as a boundary state in the tensor product of all the CFTs. To obtain the folded picture, we reflect $n-1$ of the $n$ CFTs glued at the interface so that all the CFTs are to the left of the interface; the reflection operation interchanges the left and right movers in each of these $n-1$ CFTs. In this case, the in-Hilbert space $\mathcal{H}_{\rm in}$ is the tensor product of the right moving excitations in all the CFTs and the out-Hilbert space $\mathcal{H}_{\rm out}$ is the tensor product of the left moving excitations in all the CFTs (see Fig.~\ref{Fig:setup}). The interface is then a quantum map from $\mathcal{H}_{\rm in}$ to $\mathcal{H}_{\rm out}$ in the universal sector at the linearized order.

In \cite{Chakraborty:2025dmc}, it was shown that the stringy excitations of a two-way gravitational junction can be translated, in the dual interface CFT, to a universal $\mathcal{H}_{\rm in}\to\mathcal{H}_{\rm out}$ quantum map from the space of incoming excitations to that of outgoing excitations at the interface. This map was shown to be a composition of the form $\mathcal{S}\circ\mathcal{D}$.  Here $\mathcal{D}$ is an energy redistribution of the incoming energy modes at the interface, which is determined by the normalizable stringy modes of the gravitational junction. This energy redistribution is followed by the universal scattering $\mathcal{S}$. The $\mathcal{H}_{\rm in}\to\mathcal{H}_{\rm out}$ map can also be written as  ${\mathcal{D}}\circ\mathcal{S}$, where ${\mathcal{D}}$ acts on the outgoing energy modes. 
A conformal transformation of the in or out space, parametrized by the stringy modes, can be used to realize the energy re-distribution $\mathcal{D}$.
Equivalently, the stringy excitations can also be translated to a universal $\mathcal{H}_{1}\to\mathcal{H}_{2}$ map from the Hilbert space of one CFT to that of the other.
Remarkably, for any set of values of the incoming energy modes, the stringy modes can be appropriately chosen to render the interface factorizing (perfectly reflecting) or quasi-topological (perfectly transmitting). Below, we generalize these results to the $n$-way holographic interface.

\subsection{Decoding stringy modes in terms of quantum  maps}
In holographic duality, it is crucial to understand the implications of boundary conditions. In Sec.~\ref{Sec:holoint} we have imposed Dirichlet boundary conditions on $x_s$ and $x_{d_i}$, however, we have not yet imposed any condition on $\tau_{d_i}$ and $\sigma_{d_i}$. The boundary values of $\tau_{d_i}$ and $\sigma_{d_i}$ are proportional to each other, and are determined by the energy modes $\mathcal{L}_{\pm}^{j}$ and the normalizable stringy modes $\mathcal{A}_{\omega,n}^j$. If we impose Dirichlet boundary conditions on $\tau_{d_i}$, these automatically impose the Dirichlet boundary conditions on $\sigma_{d_i}$, and fix all the normalizable stringy modes $\mathcal{A}_{\omega,n}^j$ in terms of the energy modes $\mathcal{L}_{\pm}^{j}$. However, as shown in \cite{Chakraborty:2025dmc}, this procedure realizes only a specific type of interface, and even in this case it does not reveal the universality of the scattering process which is expected \cite{Meineri:2019ycm} in a generic conformal interface. 

The crucial insight in \cite{Chakraborty:2025dmc} was that we should \emph{not} impose Dirichlet boundary conditions for $\tau_{d_i}$ and $\sigma_{d_i}$. The boundary value of $\tau_{d_i}$ introduces a relative time between the $n^{\rm th}$ and the  $(i+1)^{\rm th}$ CFT as evident from \eqref{Eq:xsxd}. As shown below, this \textit{time jump} can be undone by a conformal transformation of the $(i+1)^{\rm th}$ CFT, generalizing the method of \cite{Chakraborty:2025dmc}. Thus, by suitable $n-1$ one-sided conformal transformations we can establish continuous coordinates and metric at the interface between any pair of CFTs. 

To understand the relative time reparametrizations we first obtain $\mathbb{t}_{d_i}(\tau)$, the time shift at the interface between the $n^{\rm th}$ and $(i+1)^{\rm th}$ wire from the boundary value of $\tau_{d_i}(\sigma,\tau)$:
\begin{equation}
\lim_{\sigma\rightarrow0}\tau_{d_i}(\sigma,\tau) = \epsilon\mathbb{t}_{d_i}(\tau), \quad i= 1,\cdots,n-1.
\end{equation}
It follows from \eqref{Eq:sdtd-sol} that the boundary value of $\sigma_{d_i}$ ($\lim_{\sigma\to0} \frac{\sigma_{d_i}}{\sigma}$) is proportional to $\mathbb{t}_{\omega,d_i}$. 
Without loss of generality, we can choose the time coordinate $t_n$ of the $n^{\rm th}$ wire to be the global time. We can then express the time coordinates of the remaining $n-1$ wires in terms of $t_n$ in the form
\begin{equation}\label{Eq:timereparam}
    t_i = \mathbb{h}_i(t_n), \,\, i = 1,\hdots,n-1.
\end{equation}
We readily note from \eqref{Eq:xsxd} that
\begin{equation}\label{eq:hiexpression}
    \mathbb{h}_i(\tau)=\tau -\epsilon n\mathbb{t}_{d_{n-i}}(\tau), \ \ i=1,\cdots ,n-1.
\end{equation}
Finally, we define $\mathbb{t}_{\omega,d_i}$ using
\begin{equation}\label{Eq:t-Fd}
   \mathbb{t}_{d_i}(\tau) =\mathbb{t}_{\omega,d_i} e^{i\omega\tau}.
\end{equation}
Recall that the linear nature of the first order perturbations and the static nature of the exact zeroth order solution allow us to choose a single frequency $\omega$ without loss of generality.

Explicitly, after imposing the ingoing boundary conditions on $x_{d_i}$ and the Dirichlet boundary conditions on $x_s$ and $x_{d_i}$, we get the following expression for $\mathbb{t}_{\omega,d_i}$:
\begin{align}\label{Eq:time-repar-exps}
    \mathbb{t}_{\omega,d_i}&=\frac{i}{n\omega^3}\Bigg[\frac{\lambda}{n}\Bigg(\sum_{j=1}^{n-1}\mathcal{L}^{j}_{\omega,-}\Bigg(\mathcal{T}_{\omega}^{n}-\mathcal{T}_{\omega}^{n-i}\Bigg)+\mathcal{L}^{n-i}_{\omega,-}\left(\sum_{j=1}^{n-1}\mathcal{T}_{\omega}^{j}+\mathcal{T}_{\omega}^{n}\right) - \mathcal{L}^{n}_{\omega,-}\left(\sum_{j=1}^{n-1}\mathcal{T}_{\omega}^{j}+\mathcal{T}_{\omega}^{n-i}\right) \Bigg) \nonumber\\
    &\qquad \qquad \qquad \qquad \qquad + \frac{4}{n}a_{\omega,n}^{(i)} + \Big(\left(\mathcal{L}^{n}_{\omega,+}-\mathcal{L}^{n-i}_{\omega,+}\right) + \left(\mathcal{L}^{n}_{\omega,-}-\mathcal{L}^{n-i}_{\omega,-}\right)\Big)\Bigg],
\end{align}
directly from \eqref{Eq:sdtd-sol} using \eqref{Eq:ingoing_bc}, \eqref{Eq:param_L_modes} and \eqref{Eq:Ai_nn_Dir}. Above, we have used the rescaled normalizable modes $a_{\omega,n}^{(i)}$ which are defined as
\begin{equation}\label{Eq:a}
    a_{\omega,n}^{(i)} = \frac{i}{\sqrt{8n \pi}}{\lambda \left(n^2-\lambda ^2\right)^{3/4} \omega ^{3/2}}\mathcal{A}_{\omega,n}^i\, , \ \ i=1,\cdots ,n-1.
\end{equation}

{ In the tensionless ($\lambda\rightarrow 0$) limit, we note that the last two terms, namely
\[
\left(\mathcal{L}^{n}_{\omega,+}-\mathcal{L}^{n-i}_{\omega,+}\right) + \left(\mathcal{L}^{n}_{\omega,-}-\mathcal{L}^{n-i}_{\omega,-}\right),
\]
in the second line of \eqref{Eq:time-repar-exps} survive. However, as shown in \cite{Chakraborty:2025jtj}, the normalizable modes of $x_{d_i}$ decay as $\sigma^{-2}$ instead of $\sigma^{-3}$ near the boundary, and $\mathbb{t}_{\omega,d_i}$ is determined by the quadratic combination of these normalizable modes in the tensionless limit at the leading order in the amplitude expansion. This occurs as $\lambda$ is a singular perturbation to the Nambu-Goto-Monge-Amp\`{e}re equations followed by $x_{d_i}$. We will return to this issue in Sec. \ref{sec:bs}. For now, we ignore these quadratic terms in the tensionless limit as we discuss only the $\mathcal{O}(\epsilon)$ linear terms.}

The time shifts (jumps) above can be undone, and continuous coordinates and metric can be established across any pair of CFTs at the interface using conformal transformations on the $n-1$ wires (on all except the $n^{\rm th}$ one). Let $x^\pm_i = t_i\pm x_i$ be the lightcone coordinates of the wires. The $n-1$ conformal transformations involve the coordinate transformations
\begin{equation}\label{Eq:CTs}
    \tilde{x}^\pm_i = \mathbb{h}_i^{-1}(x^\pm_i), \,\, i = 1,\cdots,n-1,
\end{equation}
with $\mathbb{h}_i$ defined in \eqref{eq:hiexpression}, and the associated Weyl scalings that bring the metric back to the Minkowski metric. These transformations give the following new coordinates on the $n-1$ wires:
\begin{align}
    \tilde{t}_i &=\frac{1}{2}\lb\mathbb{h}_i^{-1}(t_i+x_i)+\mathbb{h}_i^{-1}(t_i-x_i)\rb, \label{Eq:newt}\\
    \tilde{x}_i &=\frac{1}{2}\lb\mathbb{h}_i^{-1}(t_i+x_i)-\mathbb{h}_i^{-1}(t_i-x_i)\rb.\label{Eq:newx}
\end{align}
It is easy to see from \eqref{Eq:newx} that the conformal transformations preserve the spatial location of the interface at $\tilde{x}_i=0$ at all times since $\tilde{x}_i(t_i, x_i =0)=0$. Furthermore, we readily see from \eqref{Eq:newt} and \eqref{Eq:timereparam} that the conformal transformations ensure that the time coordinates are continuous as we move across the interface from any CFT to another, since
\begin{equation}
    \tilde{t}_i(x_i =0,t_i)= t_n, \, i=1,2,\hdots n.
\end{equation}
Following \cite{Balasubramanian:1999re,Skenderis:2008dg,deHaro:2000vlm}, these conformal transformations can be uplifted to bulk diffeomorphisms $X_i^{\mu}(\sigma,\tau) \rightarrow \tilde{X}_i^{\mu}(\sigma,\tau)$ with $X$ and $\tilde{X}$ denoting the bulk coordinates of the $i^{\rm th}$ bulk spacetime corresponding to the $i^{\rm th}$ CFT. Since the induced metric and extrinsic curvatures of $\Sigma_{i}$ remain invariant under these bulk diffeomorphisms, we obtain an equivalent solution of the junction conditions.

However, it is easy to see that the most general conformal transformations which achieve continuous coordinates and metric across the interface for any pair of CFTs are given by
\begin{equation}\label{Eq:CTs-gen}
    \tilde{x}^\pm_i = \mathbb{k}_i^{-1}(x^\pm_i), \,\, i = 1,\cdots,n,
\end{equation}
with
\begin{align}
  \mathbb{k}_i^{-1}(x^\pm_i) = \begin{cases}
      f(\mathbb{h}_i^{-1}(x^\pm_i)), \,\, i = 1,\cdots,n-1,\\
      f(x_i^\pm), \,\, i =n,
  \end{cases}
\end{align}
where $x^\pm\to f(x^\pm)$ is a uniform conformal transformation of all the $n$-CFTs. Assuming $f$ has a well defined Fourier decomposition in terms of the frequency modes, and a perturbative expansion in $\epsilon$, we get
\begin{align}
  \mathbb{k}_i^{-1}(x^\pm_i) = \begin{cases}
      x_i^\pm+\epsilon(\mathbb t_{\omega,d_{n-i}}+f_\omega) e^{i\omega x_i^\pm}, \,\, i = 1,\cdots,n-1,\\
      x_i^\pm+\epsilon f_\omega e^{i\omega x_i^\pm} , \,\, i =n,
  \end{cases}
\end{align}
where we have used \eqref{eq:hiexpression} and \eqref{Eq:t-Fd}. As mentioned before, we are restricting to a single frequency mode without loss of generality to analyze the linear first order perturbations.

Due to the conformal transformations \eqref{Eq:CTs-gen}, the energy-momentum tensor becomes discontinuous at the interface with the following non-vanishing components for $i=1,\cdots,n$
{\begin{align}\label{Eq:Tmni}
   \widetilde{T}_{\pm\pm}^i(\tilde{x}_i^\pm) &=\frac{\pi c}{12} \mathbb{k}_i'(\tilde{x}_i^\pm)^2 T_\pm^i(\mathbb{k}_i(\tilde{x}_i)^\pm) -\frac{c}{24\pi}{\rm Sch}(\mathbb{k}_i(\tilde{x}_i^\pm),\tilde{x}_i^\pm) \nonumber\\
   &=\frac{c\epsilon}{12\pi}e^{i \omega \tilde{x}_i^\pm}\widetilde{\mathcal{L}}_{\omega,\pm}^{i}+ \mathcal{O}(\epsilon^2).
\end{align}}
The above transformations of the energy-momentum tensor are reproduced by the holographic renormalization procedure \cite{Balasubramanian:1999re,Henningson:1998gx,deHaro:2000vlm} via the bulk diffeomorphisms that uplift the conformal transformations \eqref{Eq:CTs}. We discuss the Ward identities for the energy-momentum tensor in Sec. \ref{Sec:WIs}. Since experiments are physically set up in coordinates and metric which are continuous across any pair of half-lines glued at the interface, $\widetilde{\mathcal{L}}_{\omega,\pm}$ denote the physical excitations of the CFTs. Note that the parameter $f_\omega$ determines the background state which is identical for all the identical CFTs glued at the interface, and which is generically different from the vacuum. The departures from the background gives the $\mathcal{H}_{\rm in}\rightarrow \mathcal{H}_{\rm out}$ quantum map.

In order to obtain the $\mathcal{H}_{\rm in}\rightarrow \mathcal{H}_{\rm out}$ quantum map explicitly, we note that $\widetilde{\mathcal{L}}_{\omega,-}^i$ are the physical incoming excitations at the interface and $\widetilde{\mathcal{L}}_{\omega,+}^i$ are the physical outgoing excitations. We note from \eqref{Eq:param_L_modes} that prior to the conformal transformations ${\mathcal{L}}_{\omega,+}^i$ can be parametrized in terms of ${\mathcal{L}}_{\omega,-}^i$ and the $n$ parameters $\mathcal{T}_{\omega}^i$. Additionally, $f_\omega$ (the background) and $a_{\omega,n}^i$ (the intrinsic stringy modes of the junction) determine the conformal transformations. Therefore, the physical incoming and outgoing energy modes are determined by $f_\omega$, $\mathcal{T}_{\omega}^i$ and $a_{\omega,n}^i$. 

After explicit conformal transformations \eqref{Eq:Tmni} to obtain the modes of the physical energy-momentum tensor  in each CFT in the continuous coordinates, we find that the outgoing energy modes are related to the ingoing energy modes following
\begin{equation}\label{Eq:transformamp}
    \widetilde{\mathcal{L}}_{\omega,+}^i=\frac{4}{n(n+\lambda)}\left(n a_{\omega,n}^{(n-i)}-\sum_{j=1}^{n-1} a_{\omega,n}^{(j)}\right)+\sum_{j=1}^n\mathcal{S}_{ij} \widetilde{\mathcal{L}}_{\omega,-}^j, \quad i=1,2,\cdots ,n \, ,
\end{equation}
where we define $a_{\omega,n}^{(0)}=0$. { In the tensionless limit, only the scattering matrix $\mathcal{S}$ contributes at the linearized order. As noted above, the normalizable modes of $x_{d_i}$ affect the time jumps quadratically at the leading order in the amplitude expansion in the tensionless limit. So they do not appear in the above linear relation between the physical energy modes in this limit. }

Above, the $n\times n$ matrix $\mathcal{S}$ is the generalization of the two-way scattering matrix \cite{Meineri:2019ycm,Bachas:2020yxv,Chakraborty:2025dmc} to the $n$-way interface, and is given by
\begin{equation} \label{Eq:S_gen}
    \mathcal{S} =\begin{pmatrix}
   \frac{(2-n)+\lambda}{n+\lambda} & \frac{2}{n+\lambda} & \cdots & \frac{2}{n+\lambda} \\
\frac{2}{n+\lambda} & \frac{(2-n)+\lambda}{n+\lambda} & \cdots & \frac{2}{n+\lambda} \\
\vdots & \vdots & \ddots & \vdots \\
\frac{2}{n+\lambda} & \frac{2}{n+\lambda} & \cdots & \frac{(2-n)+\lambda}{n+\lambda}
\end{pmatrix}.
\end{equation}
This scattering matrix $\mathcal{S}$ agrees with the analysis in \cite{Liu:2025khw} done in the absence of Nambu-Goto vibrations.\footnote{See \cite{Guo:2026prm} for related works.}
The physical energy modes \eqref{Eq:transformamp} also satisfy energy conservation and thus preserve the conformal boundary condition as
\begin{equation}\label{Eq:energycons}
   \sum_{i=1}^{n} \widetilde{\mathcal{L}}^{i}_{\omega,+}  =\sum_{i=1}^{n} \widetilde{\mathcal{L}}^{i}_{\omega,-}. 
\end{equation}
Remarkably, the $n+1$ parameters $f_\omega$ and $\mathcal{T}_{\omega}^i$ do not appear in the relation between the physical outgoing and ingoing modes given by \eqref{Eq:transformamp} and \eqref{Eq:S_gen}. The absence of $f_\omega$ in \eqref{Eq:transformamp} and \eqref{Eq:S_gen} establishes that the $\mathcal{H}_{\rm in}\rightarrow \mathcal{H}_{\rm out}$ quantum map is independent of the background state. We discuss the significance of the absence of $\mathcal{T}_{\omega}^i$ in \eqref{Eq:transformamp} and \eqref{Eq:S_gen} below.

The relation \eqref{Eq:transformamp} implies that the $\mathcal{H}_{\rm in}\rightarrow \mathcal{H}_{\rm out}$ quantum map is of the factorized form $ {\mathcal{D}}\circ\mathcal{S}$, which is the composition of the universal scattering matrix $S$ given by \eqref{Eq:S_gen} with an energy redistribution map ${\mathcal D}$ of the outgoing energy modes which is determined in terms of the intrinsic stringy modes $a_{\omega,n}^i$ of the gravitational junction. We note that ${\mathcal D}$ can be realized with the following conformal transformations on the out Hilbert space (in the tensor product of the left moving sectors of the $n$ CFTs)
\begin{equation}\label{Eq:fexp}
    g_+^j(x^{+}_j) = x_j^+ + 8i \epsilon\frac{e^{i \omega x_j^+}}{\omega^3n(n+\lambda)} \left(n a_{\omega,n}^{(n-j)}-\sum_{k=1}^{n-1} a_{\omega,n}^{(k)}\right).
\end{equation}

The physical energy modes \eqref{Eq:transformamp} can also be expressed as
\begin{equation}\label{Eq:transformamp2}
    \widetilde{\mathcal{L}}^{i}_{\omega,+}=\sum_{j=1}^n\mathcal{S}_{ij} \left(\widetilde{\mathcal{L}}^{j}_{\omega,-} + \frac{4}{n(n+\lambda)}\left(n a_{\omega,n}^{(n-i)}-\sum_{k=1}^{n-1} a_{\omega,n}^{(k)}\right)\right).
\end{equation}
The above implies that the  $\mathcal{H}_{\rm in} \to \mathcal{H}_{\rm out}$ map can be rewritten in the form $\mathcal{S}\circ\mathcal{D}$ where $\mathcal{D}$ redistributes energy in the in-Hilbert space. { We comment on the analogues of these one-sided conformal transformations in the tensionless limit in Sec. \ref{sec:bs}.}

Since $ \mathbb{t}_{d_i}$ should be viewed as sources (boundary values of $\tau_{d_i}$) and $ \mathbb{t}_{d_i}$ specify $a_{\omega,n}^{(n-i)}$ uniquely, we can view each stringy mode configuration as realizing a distinct interface corresponding to a specific quantum map in accordance with the tenets of the holographic duality. Note that the Dirichlet boundary conditions $\tau_{d_i}(\sigma =0) = \mathbb{t}_{d_i} =0$ imply that the stringy modes $a_{\omega,n}^{(n-i)}$ should vanish. These realize only a special case of our most general result.

We note that the disappearance of $\mathcal{T}_\omega^i$ in the relations \eqref{Eq:transformamp} and \eqref{Eq:transformamp2}, and the permutation symmetry of the scattering matrix $\mathcal{S}$ which is evident from \eqref{Eq:S_gen} imply that the universal (background independent) energy scattering occurs independently and symmetrically for an arbitrary incoming state. These features of independent and symmetric scattering for arbitrary inputs could not be assumed despite the linearity of the scattering process at the first order in the perturbation expansion. Since the gravitational problem is fundamentally non-linear, such an assumption at the linear level can potentially affect the results at higher orders. We have achieved the demonstration of independent and symmetric scattering for arbitrary inputs as the generic asymmetric $\mathcal{T}_\omega^i$ (which represent transmission coefficients prior to the transformation to the physical conformal frame) disappear in the quantum map relating the physical incoming and outgoing energy modes.

Several comments are in order. Firstly, from \eqref{Eq:S_gen} we note that the reflection coefficient $\mathcal{R}$ in any one of the CFTs in the multi-interface is
\begin{equation}\label{Eq:Rineq}
    \mathcal{R} = \frac{2-n +\lambda}{n+\lambda} \geq \frac{2-n}{n}
\end{equation}
for $0\leq \lambda < n$. This is consistent with reflection positivity \cite{Billo:2016cpy,Meineri:2019ycm} which requires that
\begin{equation}
    \mathcal{R} \geq \frac{c_L - c_R}{c_L+c_R}
\end{equation}
when energy is transmitted from CFT$_L$ with central charge $c_L$ to CFT$_R$ with central charge $c_R$.\footnote{We thank Marco Meineri for discussions on reflection positivity and ANEC bounds.} In this specific case, $c_R = (n-1)c_L$ as the energy transmission occurs from one of the wires to the remaining $n-1$ wires with each wire described by a CFT with identical central charge. Therefore, $\dfrac{c_L - c_R}{c_L+c_R} = \dfrac{2-n}{n}$ which is precisely the lower bound in \eqref{Eq:Rineq} that is saturated when $\lambda = 0$. 

{ Particularly, the $\mathcal{S}$-matrix given by \eqref{Eq:S_gen} is explicitly}
\begin{equation} \label{Eq:S_gen_t0}
    \mathcal{S} =\begin{pmatrix}
   \frac{(2-n)}{n} & \frac{2}{n} & \cdots & \frac{2}{n} \\
\frac{2}{n} & \frac{(2-n)}{n} & \cdots & \frac{2}{n} \\
\vdots & \vdots & \ddots & \vdots \\
\frac{2}{n} & \frac{2}{n} & \cdots & \frac{(2-n)}{n}
\end{pmatrix},
\end{equation}
{ in the tensionless limit, featuring the maximally negative energy reflection coefficient allowed by reflection positivity when all wires are described by CFTs with the same central charge. This $\mathcal{S}$-matrix is remarkably of the same form as Grover diffusion operator used in quantum walks \cite{Childs:2004tgh}, and has also appeared in the context of scale-invariant $S$-matrices for particle scattering in free quantum fields on star graphs \cite{Bellazzini:2006jb} and in junctions of anyonic Luttinger wires  at scale-invariant fixed points \cite{Bellazzini:2008fu} indicating other physical contexts in which such a negative reflection coefficient appears. It would be interesting to understand the general conditions under which such a scattering matrix can appear in a scale-invariant interface.}

Secondly, for the full range of values of $\lambda$ given by $0\leq\lambda< n$, the eigenvalues of $\mathcal{S}^T\mathcal{S}$ are between $0$ and $1$ indicating that the scattering process satisfies unitarity bounds.

Since the scattering matrix $\mathcal{S}$ is frequency independent, $\mathcal{S}$ also applies for the expectation values of the ANEC operators which are  $\widetilde{\mathcal{L}}^i_{\omega =0, \pm}$, the incoming and outgoing total energy fluxes. For the case of a two-way interface between two holographic CFTs with different temperatures {(given by the zero frequency modes above)}, it has been shown in \cite{Bachas:2021tnp} that a steady state heat current develops precisely as expected from the scattering matrix $\mathcal{S}$. It would be of interest to generalize this result for the steady state heat currents in the multiway junction.

The ANEC refers to the averaged null energy condition which implies that \cite{Quella:2006de,Meineri:2019ycm} 
\begin{equation}
0 \leq \mathcal{R}\leq 1
\end{equation}
when energy is transmitted from one of the CFTs to the remaining $n-1$ CFTs. The lower ANEC bound is violated when $0\leq \lambda < n - 2$. This motivates us to re-examine if the ANEC holds in the presence of an interface and/or boundaries especially as the interface and/or boundaries break the Poincar\'{e} invariance explicitly and the null geodesics where the ANEC operators are measured touch the interface/boundaries unlike what have been assumed in the proofs of the ANEC \cite{Klinkhammer:1991ki,Kelly:2014mra,Faulkner:2016mzt,Hartman:2016lgu,Kravchuk:2018htv}.

\subsection{Ward identities}\label{Sec:WIs}

We can study the Ward identities for the multi-way junction following \cite{Chakraborty:2025jtj}. To formulate the Ward identities, we consider a bi-partition in which the $i^{\rm th}$ wire is on the left of the interface and the other $n-1$ wires are to the right of the interface. For the latter, we need to apply reflection about the interface at $\tilde{x}=0$ on the second set of $n-1$ wires which exchanges the left and right movers in each of these CFTs (note that we have used the folded picture so far). Thus the interface glues CFT$_L$, which is CFT$_i$ with the tensor product of the remaining $\overline{\rm CFT}_j$ (with $j\neq i$) where the overline denotes the action of reflection. As a result, the non-vanishing energy-momentum tensor components across the interface are
    \begin{align}
    &\widetilde{T}_{++}(\tilde{t},\tilde{x}) =\Theta(-\tilde{x}) \widetilde{T}_{++}^i(\tilde{t},\tilde{x})+
         \Theta(\tilde{x})\sum_{j\neq i}\widetilde{T}^j_{--}(\tilde{t},\tilde{x})\\
    &\widetilde{T}_{--}(\tilde{t},\tilde{x}) = 
         \Theta(-\tilde{x}) \widetilde{T}_{--}^i(\tilde{t},\tilde{x})+
         \Theta(\tilde{x})\sum_{j\neq i}\widetilde{T}^j_{++}(\tilde{t},\tilde{x}).
\end{align}

The Ward identities are \cite{Chakraborty:2025dmc,Chakraborty:2025jtj}
\begin{align}
    &\partial_{\tilde{t}}\widetilde{T}^{\tilde{t}\tilde{t}}(\tilde{t},\tilde{x})+\partial_{\tilde{x}}\widetilde{T}^{\tilde{x}\tilde{t}}(\tilde{t},\tilde{x}) =0,\label{Eq:WI1}\\
    &\partial_{\tilde{t}}\widetilde{T}^{\tilde{t}\tilde{x}}(\tilde{t},\tilde{x})+\partial_{\tilde{x}}\widetilde{T}^{\tilde{x}\tilde{x}}(\tilde{t},\tilde{x})={\delta(\tilde{x})}q(\tilde{t}).\label{Eq:WI2}
\end{align}
    The disappearance of the source in \eqref{Eq:WI1} is a consequence of the conformal boundary condition \eqref{Eq:energycons}. Note that the discontinuity in $T^{xt} \propto T_{++}-T_{--}$ at the junction, which gives the source in \eqref{Eq:WI1}, is proportional to
    \begin{equation}
        \delta(\tilde{x}) \left(\sum_{j=1}^n\widetilde{T}^j_{++}- \sum_{j=1}^n\widetilde{T}^j_{--}\right),
    \end{equation}
    which vanishes due to \eqref{Eq:energycons}. The source appearing in the second Ward identity \eqref{Eq:WI2} is \cite{Chakraborty:2025jtj}
\begin{equation}
    q^i(\tilde{t})=\sum_{j\neq i}\lb\widetilde{T}^j_{++}+\widetilde{T}^j_{--}\rb(\tilde{t},\tilde{x}=0) -\lb\widetilde{T}^i_{++}+\widetilde{T}^i_{--}\rb(\tilde{t},\tilde{x}=0).
\end{equation}
For a two-way junction, the source of the second Ward identity \cite{Chakraborty:2025dmc} can be interpreted as the expectation value of a displacement operator, which quantifies the energy cost of an infinitesimal displacement of the interface \cite{Bianchi:2015liz}. The source above, for the multi-way interface, can be interpreted as the expectation value of a generalized displacement operator $D_i$ that generates the displacement of the $i^{\rm th}$ wire away from the interface.

Using the energy-momentum tensors \eqref{Eq:Tmni}, we see that the source is
\begin{equation}\label{Eq:qi-1}
    q^i(\tilde{t})=\frac{c\epsilon e^{i\omega\tilde{t}}}{12\pi} \lb\sum_{j\neq i}(\widetilde{\mathcal{L}}^{j}_{\omega,+}+\widetilde{\mathcal{L}}^{j}_{\omega,-}) -(\widetilde{\mathcal{L}}^{i}_{\omega,+}+\widetilde{\mathcal{L}}^{i}_{\omega,-})\rb.
\end{equation}
Using energy conservation \eqref{Eq:energycons} and \eqref{Eq:transformamp}, the above source can be expressed as
\begin{equation}\label{Eq:WIsrc}
    q^i(\tilde{t})=\frac{c\epsilon e^{i\omega\tilde{t}}}{12\pi} \Bigg[ \lb\widetilde{\mathcal{L}}^{n}_{\omega,+}+\widetilde{\mathcal{L}}^{n}_{\omega,-}-\sum_{j=1}^n\widetilde{\mathcal{L}}^{j}_{\omega,-}\rb +\frac{4}{n+\lambda}\lb a_{\omega,n}^{(n-i)} - \frac{\lambda}{2}\lb\widetilde{\mathcal{L}}^{n}_{\omega,-}-\widetilde{\mathcal{L}}^{i}_{\omega,-}\rb\rb\Bigg].
\end{equation}
Note that $q^i(\tilde{t})$, the expectation values of the displacement operators $D_i$ can be expressed as a linear combination of only the incoming modes $\widetilde{\mathcal{L}}^{j}_{\omega,-},\, j=1,\cdots, n $, and the stringy modes $a_{\omega,n}^{(i)}$, using \eqref{Eq:transformamp} for $\widetilde{\mathcal{L}}^{n}_{\omega,+}$. Equivalently, the stringy modes $a_{\omega,n}^{(i)}$ can be reconstructed from $q^i(\tilde{t})$, the expectation values of the displacement operators $D_i$ and incoming energy modes $\widetilde{\mathcal{L}}^{j}_{\omega,-}$.

{ Clearly, we can formulate other versions of these Ward identities by bi-partitioning the $n$ wires into two sets, each containing more than one wire, in a similar manner.}

\subsection{Tuning the quantum maps using stringy modes}
Remarkably, as we show below, the Nambu-Goto modes can be chosen to realize the multi-way interface. Here we study two special cases, namely the pseudo-factorizing limit in which we get purely reflexive behavior and the pseudo-topological limit in which the energy scattering matrix becomes an involution.

As can be seen from \eqref{Eq:transformamp}, the multi-way interface is pseudo-factorizing (perfectly reflecting), that is 
\begin{equation}\label{Eq:pr}
    \widetilde{\mathcal{L}}^{i}_{\omega,+}=\widetilde{\mathcal{L}}^{i}_{\omega,-}\, , \, \forall = \,i=1,2,\cdots, n\, ,
\end{equation}
when
\begin{equation}\label{Eq:an-special-1}
    a_{\omega,n}^{(i)} = \frac{n}{2}\Bigg(\widetilde{\mathcal{L}}^{n-i}_{\omega,-}-\widetilde{\mathcal{L}}^{n}_{\omega,-}\Bigg).
\end{equation}
Since $a_{\omega,n}^{(i)}$ determines a specific interface (quantum map), we note that the pseudo-factorizing behavior is realized only when the incoming energy modes satisfy the condition \eqref{Eq:an-special-1}, and not for arbitrary values of the incoming energy modes.

A natural way to generalize the topological interface is to consider the vanishing of maximally possible number of expectation values of the displacement operators ($q^i$) for $n$ bi-partitions of the wires into sets containing one and $n-1$ wires as discussed in the previous section.  For $n>2$, it is easy to see that setting all $q^i$ to zero implies that the total incoming and thus the total outgoing energy flux should both vanish. Let
\begin{equation}
    \mathcal{E}_{tot} = \sum_i \widetilde{\mathcal{L}}^{i}_{\omega,+} , \quad \mathcal{X}_i =  \widetilde{\mathcal{L}}^{i}_{\omega,+}+  \widetilde{\mathcal{L}}^{i}_{\omega,-}
\end{equation}
Since $\sum_i \widetilde{\mathcal{L}}^{i}_{\omega,+} = \sum_i \widetilde{\mathcal{L}}^{i}_{\omega,-}$ by energy conservation, we obtain
\begin{equation}\label{Eq:sum-1}
    \sum_i\mathcal{X}_i = 2\mathcal{E}_{tot}.
\end{equation}
From \eqref{Eq:qi-1} and \eqref{Eq:transformamp} it follows that
\begin{equation}
    q^i\propto \mathcal{E}_{tot} - \mathcal{X}_i.
\end{equation}
So $q^i = 0$ would imply $\mathcal{X}_i =\mathcal{E}_{tot}$, and therefore. 
\begin{equation}\label{Eq:sum-2}
    \sum_i\mathcal{X}_i = n\mathcal{E}_{tot}.
\end{equation}
Comparing with \eqref{Eq:sum-1}, we obtain
\begin{equation}
    (n-2)\mathcal{E}_{tot} =0
\end{equation}
if all $q^i$ vanish. Therefore, for $n>2$, the expectation values of all displacement operators can vanish only if we restrict the total energy input to zero. 

If we do not want to restrict the ingoing energy modes, then we can impose maximally $n-1$ of the displacement operators to vanish. Without loss of generality, let us set $q^i =0$ for $i = 1, 2, \cdots, n-1$ for $n\geq 3$. From \eqref{Eq:WIsrc}, it is easy to see that this condition can be achieved by fine-tuning $a_{\omega,n}^{(i)}$ so that
\begin{equation}\label{Eq:an-ptop}
    a^{(n-i)}_{\omega,n}
 =
 \frac{\lambda}{2}
 \left(\widetilde{\mathcal{L}}^{n}_{\omega,-}  - \widetilde{\mathcal{L}}^{i}_{\omega,-} \right)
 +\frac{(n+\lambda)(n-2)}{4}\mathcal{E}_{tot}
 \qquad i=1,\ldots,n-1.
\end{equation}
Then we obtain from \eqref{Eq:transformamp}
\begin{align}
    \begin{pmatrix}
\widetilde{\mathcal{L}}^{1}_{\omega,+}\\
\widetilde{\mathcal{L}}^{2}_{\omega,+}\\
\vdots\\
\widetilde{\mathcal{L}}^{n-1}_{\omega,+}\\
\widetilde{\mathcal{L}}^{n}_{\omega,+}
\end{pmatrix}
=
\underbrace{
\begin{pmatrix}
0 & 1 & 1 & \cdots & 1 & 1\\
1 & 0 & 1 & \cdots & 1 & 1\\
\vdots & \vdots & \ddots & \ddots & \vdots & \vdots\\
1 & 1 & \cdots & 0 & 1 & 1\\
1 & 1 & \cdots & 1 & 0 & 1\\
3-n & 3-n & \cdots & 3-n & 3-n & 2-n
\end{pmatrix}
}_{\displaystyle \mathsf S_{1\mid(n-1)}}
\begin{pmatrix}
\widetilde{\mathcal{L}}^{1}_{\omega,-}\\
\widetilde{\mathcal{L}}^{2}_{\omega,-}\\
\vdots\\
\widetilde{\mathcal{L}}^{n-1}_{\omega,-}\\
\widetilde{\mathcal{L}}^{n}_{\omega,-}
\end{pmatrix}.
\end{align}
This pseudo-topological energy scattering matrix $\mathsf S_{1\mid(n-1)}$ for the incoming energy modes which satisfy \eqref{Eq:an-ptop} not only conserves the total energy but is remarkably also formally an involution as it satisfies
\begin{equation}
    \mathsf S_{1\mid(n-1)}^2= I_n.
\end{equation}
For $n=2$, the topological junction gives perfect energy transmission so that the energy scattering matrix is just $I_2$. For $n\geq 3$, we see that the pseudo-topological condition stating that maximal number of expectation values of the displacement operators vanish implies that the energy scattering matrix is an involution. While $q^i$ vanishes for $1\leq i \leq n-1$, we can readily see that $q^n \propto (n-2)\mathcal{E}_{tot}$.

\subsection{Boundary state interpretation and beyond linearized perturbations}\label{sec:bs}

Our results suggest that the holographic interface dual to the solution of the gravitational junction corresponding to a specific solution of the intrinsic Nambu-Goto modes is a boundary state $\ket{\mathcal{B}}$ of the tensor product of the $n$ CFTs in the folded picture satisfying:
\begin{align}\label{Eq:int}
    \sum_i\left( \widetilde{L}_{n,-}^i - \widetilde{L}_{-n,+}^i \right)\ket{\mathcal{B}} = 0,
\end{align}
where $\widetilde{L}_{n,\pm}^i$ are the generators of the Virasoro algebra of the $i^{\rm th}$CFT after automorphisms which are determined by the Nambu-Goto modes. The above follows from \eqref{Eq:energycons} which must be satisfied by the physical energy modes $\widetilde{\mathcal{L}}_{n,\pm}^i$ implying the conformal boundary condition in terms of the Virasoro algebra of each of the $n$-CFTs post automorphisms determined by the Nambu-Goto modes $a_{\omega,n}^{(i)}$. We leave the important task of determining $\ket{\mathcal{B}}$ explicitly for a given choice of the Nambu-Goto modes $a_{\omega,n}^{(i)}$ for the future. 

It is pertinent to comment that \eqref{Eq:int} is compatible with the non-linearity of the gravitational problem provided we consider automorphisms which are more general than conformal transformations. Consider the automorphism 
\[
L_n \rightarrow \widetilde{L}_n= \mathcal{V}^{-1} L_n \mathcal{V}
\]
with
\[
\mathcal{V} = \exp\left(-\sum_m \alpha_m L_m - \sum_{i=2}^\infty\sum_{\{p_1,p_2,\cdots ,p_i\}}\gamma_{p_1p_2\cdots p_i}L_{p_1}L_{p_2}\cdots L_{p_i}\right).
\]
Clearly $\widetilde{L}_n$ is an automorphism of the Virasoro algebra. In absence of the higher order terms $\gamma_{p_1p_2\cdots p_i}$, $\alpha_m$ implement conformal transformations for which $\{\widetilde{L}_n\}$ are related to $\{L_n\}$ linearly. The general automorphisms are non-linear transformations of the Virasoro generators, e.g. $L_{-4}$ can mix with $L_{-2}^2$. Such a non-linear mixing is needed to be compatible with the higher order gravitational perturbations which mix energy modes of different frequencies. The higher order coefficients $\{\gamma_{p_1p_2\cdots p_i}\}$ should also be determined by the Nambu-Goto modes $\{a_{\omega,n}^{(i)}\}$ of the gravitational junction just like ${\alpha_m}$ which implement the conformal transformations.\footnote{Note $\{\gamma_{p_1p_2\cdots p_i}\}$ do not depend on the energy modes simply by construction, and so there is no state-dependence in the definition \eqref{Eq:int} of the boundary state. If we are expanding the physical energy modes as a polynomial of the energy modes (prior to automorphisms), then the coefficients of this polynomial depend on $\{a_{\omega,n}^{(i)}\}$ only. When we rewrite this relation in the operator form $\widetilde{L}_n= \mathcal{V}^{-1} L_n \mathcal{V}$, the coefficients $\{\gamma_{p_1p_2\cdots p_i}\}$ similarly do not depend on the incoming energy modes.} In the future, we intend to explicitly determine how the automorphisms of the Virasoro algebra are determined by the Nambu-Goto modes at higher orders in the perturbative expansion.

Recently, in \cite{Banerjee:2025zuw} the full non-linear problem has been analyzed in a special case of a two-way junction when both CFTs are glued at the dual interface with the same background temperature with absence of energy modes on the left CFT. The global solution can be understood by resumming the perturbative expansion, using techniques developed in \cite{Banerjee:2023djb,Mitra:2024zfy}. This shows that the wavepackets corresponding to the stringy excitations are born out of initial conditions (gravitational memory) at past null infinity. These are then incident on the interface where they are perfectly reflected, without distortion, to future null infinity. This perfect reflection process is manifestly causal. As hinted in \cite{Chakraborty:2025jtj}, this particular simpler setup can be readily generalized to the multiway junction. Particularly, the bulk solutions admit rigid parameters, which we have set to zero in the present paper. A non-linear analysis would reveal the role played by these parameters in the multi-way conformal interface generalizing \cite{Banerjee:2025zuw}. 

{ Furthermore, in the tensionless limit, the decoding of the stringy modes in terms of holographic quantum maps at the dual interface requires the above boundary state interpretation explicitly as the time jumps across the interface are quadratic in the amplitudes of the stringy modes even at the leading order \cite{Chakraborty:2025jtj}. The undoing of the time jumps would realize automorphisms of the Virasoro algebra which are not just conformal transformations (adjoint actions). It would be interesting to see this explicitly.}

\section{Summary and future directions}\label{Sec:conclusions}

In this paper, we have shown that the multi-way gravitational junction, including its stringy excitations, can be translated to quantum maps in the dual multi-way conformal interface. We have explicitly worked out the quantum maps corresponding to a junction with specific stringy excitations at the linearized order in the gravitational perturbation about the static junction. We have shown that the quantum maps are universal in the sense that they do not depend on the choice of the (generic inhomogenous) background state. { These generalize the results we have obtained earlier in \cite{Chakraborty:2025dmc, Banerjee:2025} for the two-way gravitational junction.}

When the stringy modes are switched off, the quantum maps reduce to a universal scattering of the modes that are incident on the interface. In the presence of the stringy modes, the quantum map is a composition of the universal scattering with an energy re-distribution of the in or out spaces. This energy re-distribution is a consequence of conformal transformations on $n-1$ out of the $n$ wires that are parametrized by the stringy modes. The conformal boundary condition (energy conservation between incoming and outgoing energy modes) is preserved even in the presence of the stringy excitations of the junction. This is reflected by the vanishing of the source for the energy conservation Ward identity. Furthermore, we have demonstrated that the source for the momentum conservation  Ward identity, which is the expectation value of a generalized displacement operator, depends on the stringy excitations of the junction. We have also shown that the multi-way interface dual to the holographic junction is a tunable energy transmitter.

{ One novel feature for $n\geq 3$ is that the universal energy scattering matrix can have a negative reflection coefficient which saturates the lower bound obtained from reflection positivity in the tensionless limit. The general energy scattering matrix is unitary when the static configuration is permutation-symmetric. For $n\geq 3$, asymmetric static configurations describe non-unitary energy scattering as discussed in Appendix \ref{Sec:assymetricsoln}. Non-unitary conformal binary interfaces in unitary CFTs have recently been studied in the literature \cite{Tang:2026ccf}. It would be interesting to see if explicit constructions of non-unitary conformal multi-interfaces can reproduce holographic results.}

It is important to extend our results to the full non-linear setting. We have discussed how our results imply a boundary state formulation \eqref{Eq:int} of the general holographic interface which includes automorphisms of the Virasoro algebras of the CFTs glued at the interface. It is pertinent to understand whether the interpretation indeed holds to all orders in perturbation theory. Furthermore, it is of interest to explicitly determine the boundary states corresponding to the general holographic interface dual to the gravitational junction with its intrinsic excitations. In this respect, we need to understand (i) how to treat the zero modes non-perturbatively and include the steady state heat current found in \cite{Bachas:2021tnp}, and (ii) analyze especially the tensionless limit in which matter like vibrations arise in the bulk out of pure gravity. These developments could lead to the understanding of the general construction of tunable energy transmitters using quantum critical systems, and a more profound understanding of semi-classical gravity.

We note that the entanglement structure of the dual field theory plays an important role in understanding emergence of the bulk spacetime. This is therefore an essential first step towards the reconstruction of extended gravitational objects in terms of the boundary field theory. Performing a similar analysis of the entanglement entropy in the multi-way setting would reveal further aspects of how extended objects in the bulk are encoded in the dual field theory. We expect such a calculation to be tractable using the holographic entanglement entropy prescription \cite{RT,HRT} and the techniques developed in \cite{Kibe:2021qjy,Banerjee:2022dgv,Kibe:2024icu}, which were used in \cite{Banerjee:2025zuw}. 

In fact, it has been shown in \cite{Banerjee:2025zuw} that the entanglement entropy of a spacelike interval straddling the interface deciphers the stringy modes, even in this tension-less limit of the two-way gravitational junction { which is non-trivial in the presence of rigid parameters (that realize worldsheet isometries or one-sided spacetime isometries which preserve the interface at the boundary.)}. { It is to be noted that the tensionless limit is non-trivial even in the absence of rigid parameters for $n\geq 3$ due to the presence of the dynamical Monge-Amp\`{e}re modes of the gravitational junction which survive in this tensionless limit. Therefore, it would be of interest to understand how entanglement in the dual interface detects the emergence of matter-like vibrations from pure gravity.}

\begin{acknowledgments}
We thank Costas Bachas, Marco Meineri, Mihail Mintchev and Baishali Roy for valuable discussions and comments on the manuscript. AC, AM and MM acknowledge support from FONDECYT postdoctoral grant no. 3230222, FONDECYT regular grant no. 1240955 and ``Doctorado Nacional'' grant no. 21250596 of La Agencia Nacional de Investigaci\'{o}n y Desarrollo (ANID), Chile, respectively. TK is supported by a Simons Foundation’s fellowship through the Targeted Grant to Instituto Balseiro. AM gratefully acknowledges the hospitality of LPENS, where a substantial part of this work was carried out during his tenure as a CNRS invited professor.
\end{acknowledgments}

\begin{appendix}

\section{Analysis of the asymmetric static solutions of the junction conditions} \label{Sec:assymetricsoln}
In addition to the symmetric solution~\eqref{Eq:order0soln}, which exists for $0 <\lambda<n$, we find permutation asymmetric solutions to the junction conditions when $0<\lambda<n-2$. These solutions are unphysical in the sense that the $\mathcal{S}$ matrix obtained is non-unitary. Below we will provide details of one of these solutions for a $3$-way junction. Similar results hold for $n>3$ as well. 

One of the asymmetric solutions to the junction conditions, for $n=3$, at $\mathcal{O}(\epsilon^0)$ is
\begin{equation} \label{Eq:asym_sol}
    \tau_d =0\,, \quad \sigma_d=0\,, \quad x_d=2x_s\,, \quad x_s=\frac{\sigma \tilde{p}_{\lambda}}{3}\,,
\end{equation}
where $\tilde{p}_{\lambda}=\frac{\lambda}{\sqrt{1-\lambda^2}}$ and $0 < \lambda < 1$. At $\mathcal{O}(\epsilon)$, the transverse coordinates $x_{d_1}$ and $x_{d_2}$ satisfy the linearized Nambu-Goto equations coupled with the source terms, as
 \begin{align} \label{Eq:NG-asym}
     6\sigma\tilde{p}_{\lambda}^2\ddot{x}_{d_1} + 6\lambda^2 \lb 2 x^\prime_{d_1} - \sigma x^{\prime\prime}_{d_1}\rb &= i\omega\lambda^2 \sigma^3 e^{i\omega \tau} \lb (\mathcal{L}_{\omega,+}^2+\mathcal{L}_{\omega,-}^3)e^{-i\tilde{p}_{\lambda}\omega\sigma} - (\mathcal{L}_{\omega,-}^2+\mathcal{L}_{\omega,+}^3)e^{i\tilde{p}_{\lambda}\omega\sigma} \rb, \notag \\
     6\sigma\tilde{p}_{\lambda}^2\ddot{x}_{d_2} + 6\lambda^2 \lb 2 x^\prime_{d_2} - \sigma x^{\prime\prime}_{d_2}\rb &= i\omega\lambda^2 \sigma^3 e^{i\omega \tau} \lb (\mathcal{L}_{\omega,+}^1-\mathcal{L}_{\omega,+}^3)e^{i\tilde{p}_{\lambda}\omega\sigma} - (\mathcal{L}_{\omega,-}^1-\mathcal{L}_{\omega,-}^3)e^{-i\tilde{p}_{\lambda}\omega\sigma} \rb, 
 \end{align}
 where dot and prime denote $\partial_{\tau}$ and $\partial_{\sigma}$ respectively. We get the following solutions to the equations \eqref{Eq:NG-asym}
\begin{align}
    &x_{d_1}=\frac{e^{i\omega \tau}}{6\omega^3}\Bigg[\sqrt{-\frac{72\lambda^3\omega^3}{\pi\tilde{p}^3_{\lambda}}}\left(\sin \left(\frac{\tilde{p}_{\lambda}\omega\sigma}{\lambda}\right) \left(\mathcal{A}_{\omega,1} + \frac{\tilde{p}_{\lambda}\omega\sigma}{\lambda} \mathcal{A}_{\omega,2} \right) + \cos \left(\frac{\tilde{p}_{\lambda}\omega\sigma}{\lambda}\right) \left(\mathcal{A}_{\omega,2} - \frac{\tilde{p}_{\lambda}\omega\sigma}{\lambda} \mathcal{A}_{\omega,1} \right)\right)\nonumber\\
    &+ (\mathcal{L}_{\omega,+}^3+\mathcal{L}_{\omega,-}^2) e^{i\tilde{p}_{\lambda}\omega\sigma} \left(2\tilde{p}_{\lambda} \omega\sigma + i \left(2+\omega ^2\sigma^2\right) \right) + (\mathcal{L}_{\omega,-}^3+\mathcal{L}_{\omega,+}^2) e^{-i\tilde{p}_{\lambda}\omega\sigma} \left(2\tilde{p}_{\lambda} \omega\sigma - i \left(2+\omega ^2\sigma^2\right) \right)\Bigg], \notag \\
    &x_{d_2}=\frac{e^{i\omega \tau}}{6\omega^3}\Bigg[\sqrt{-\frac{72\lambda^3\omega^3}{\pi\tilde{p}^3_{\lambda}}}\left(\sin \left(\frac{\tilde{p}_{\lambda}\omega\sigma}{\lambda}\right) \left(\mathcal{B}_{\omega,1} + \frac{\tilde{p}_{\lambda}\omega\sigma}{\lambda} \mathcal{B}_{\omega,2} \right) + \cos \left(\frac{\tilde{p}_{\lambda}\omega\sigma}{\lambda}\right) \left(\mathcal{B}_{\omega,2} - \frac{\tilde{p}_{\lambda}\omega\sigma}{\lambda} \mathcal{B}_{\omega,1} \right)\right)\nonumber\\
    &+ (\mathcal{L}_{\omega,+}^3-\mathcal{L}_{\omega,+}^1) e^{i\tilde{p}_{\lambda}\omega\sigma} \left(2\tilde{p}_{\lambda} \omega\sigma + i \left(2+\omega ^2\sigma^2\right) \right) + (\mathcal{L}_{\omega,-}^3-\mathcal{L}_{\omega,-}^1) e^{-i\tilde{p}_{\lambda}\omega\sigma} \left(2\tilde{p}_{\lambda} \omega\sigma - i \left(2+\omega ^2\sigma^2\right) \right)\Bigg],
\end{align}ove
where the first lines are the solutions of the homogeneous part of \eqref{Eq:NG-asym}. Imposing ingoing boundary conditions we get
\begin{equation}
    \mathcal{A}_{\omega,1}=\mathcal{A}_{\omega,nn}+\mathcal{A}_{\omega,n}\,, \ \mathcal{A}_{\omega,2}=i \mathcal{A}_{\omega,nn}\, \ \ {\rm and} \ \ \mathcal{B}_{\omega,1}=\mathcal{B}_{\omega,nn}+\mathcal{B}_{\omega,n}\,, \ \mathcal{B}_{\omega,2}=i \mathcal{B}_{\omega,nn},
\end{equation}
where $\mathcal{A}_{\omega,nn},\mathcal{B}_{\omega,nn}$ are the non-normalizable modes, and $\mathcal{A}_{\omega,n},\mathcal{B}_{\omega,n}$ are the normalizable stringy modes, and can be determined by initial and boundary conditions as before. From the extrinsic curvature discontinuity we get the solution for $x_s$:
\begin{equation} \label{Eq:xs-asym}
    x_s=\frac{\sigma \tilde{p}_{\lambda}}{3} + \epsilon x_s^{(1)},
\end{equation}
with
\begin{align}
x_s^{(1)} &= \frac{e^{i\omega\tau}}{6\omega^3}\Bigg[
e^{i\tilde{p}_{\lambda}\omega\sigma} \lb \mathcal{L}_{\omega,+}^1 + \mathcal{L}_{\omega,+}^3 - \mathcal{L}_{\omega,-}^2 \rb
\lb 2 \tilde{p}_{\lambda} \omega \sigma + i\lb 2 + \omega^2 \sigma^2 \rb \rb  \notag \\
&\qquad \qquad \qquad \qquad + \, e^{-i\tilde{p}_{\lambda}\omega\sigma} \lb \mathcal{L}_{\omega,-}^1 + \mathcal{L}_{\omega,-}^3 - \mathcal{L}_{\omega,+}^2 \rb 
\lb 2 \tilde{p}_{\lambda} \omega \sigma - i\lb 2 + \omega^2 \sigma^2 \rb \rb
\Bigg].
\end{align}
Imposing Dirichlet boundary condition on $x_s$, namely, $\lim_{\sigma\to0} x_s=0$, we get
\begin{equation}
    \mathcal{L}_{\omega,+}^1 + \mathcal{L}_{\omega,+}^2 + \mathcal{L}_{\omega,+}^3 = \mathcal{L}_{\omega,-}^1 + \mathcal{L}_{\omega,-}^2 + \mathcal{L}_{\omega,-}^3 \,,
\end{equation}
which is nothing but the energy conservation \eqref{Eq:energy_con} we found for the symmetric solution. Hence, we can solve the above by using the same parametrization as \eqref{Eq:param_L_modes}, namely
\begin{align}
\mathcal{L}_{\omega,+}^3 &= \mathcal{T}_\omega^3 \mathcal{L}_{\omega,-}^1 + \mathcal{T}_\omega^3 \mathcal{L}_{\omega,-}^2 + (1 - \mathcal{T}_\omega^1 - \mathcal{T}_\omega^2) \mathcal{L}_{\omega,-}^3 \ \,, \\
\mathcal{L}_{\omega,+}^2 &= \mathcal{T}_\omega^2 \mathcal{L}_{\omega,-}^1 + (1 - \mathcal{T}_\omega^3 - \mathcal{T}_\omega^1) \mathcal{L}_{\omega,-}^2 + \mathcal{T}_\omega^2 \mathcal{L}_{\omega,-}^3 \ \,, \\
\mathcal{L}_{\omega,+}^1 &= (1 - \mathcal{T}_\omega^2 - \mathcal{T}_\omega^3) \mathcal{L}_{\omega,-}^1 + \mathcal{T}_\omega^1 \mathcal{L}_{\omega,-}^2 + \mathcal{T}_\omega^1 \mathcal{L}_{\omega,-}^3 \ \,,
\end{align}
where $\mathcal{T}_\omega^1,\mathcal{T}_\omega^2,\mathcal{T}_\omega^3$ are arbitrary coefficients. Using the Dirichlet boundary condition on $x_{d_i}$, $\lim_{\sigma\to0} x_{d_i}=0$, we solve for $\mathcal{A}_{\omega,nn}$ and $\mathcal{B}_{\omega,nn}$ in terms of these coefficients and plane wave amplitudes:
\begin{align}
    \mathcal{A}_{\omega,nn} = \sqrt{\frac{\pi}{2}}\frac{\mathcal{L}_{\omega,-}^1(\mathcal{T}_\omega^3-\mathcal{T}_\omega^2)+\mathcal{L}_{\omega,-}^2(\mathcal{T}_\omega^1+2\mathcal{T}_\omega^3)-\mathcal{L}_{\omega,-}^3(\mathcal{T}_\omega^1+2\mathcal{T}_\omega^2)}{3(1-\lambda^2)^{3/4}\omega^{3/2}}, \notag \\
    \mathcal{B}_{\omega,nn} = \sqrt{\frac{\pi}{2}}\frac{\mathcal{L}_{\omega,-}^2(\mathcal{T}_\omega^3-\mathcal{T}_\omega^1)+\mathcal{L}_{\omega,-}^1(\mathcal{T}_\omega^2+2\mathcal{T}_\omega^3)-\mathcal{L}_{\omega,-}^3(\mathcal{T}_\omega^2+2\mathcal{T}_\omega^1)}{3(1-\lambda^2)^{3/4}\omega^{3/2}}.
\end{align}
Furthermore, using the analysis in Sec.~\ref{Sec:holqmap} we have the following set of relative conformal transformations to undo the time reparameterizations 
\begin{equation}
    \mathbb{h}_1(\tau)=\tau - \epsilon\, 3\mathbb{t}_{\omega,d_2}(\tau), \ \ \mathbb{h}_2(\tau)=\tau - \epsilon\, 3\mathbb{t}_{\omega,d_1}(\tau),
\end{equation}
where,
\begin{align} \label{Eq:time_rep_asym}
    \mathbb{t}_{\omega,d_1} &= \frac{ie^{i\omega\tau}}{3\omega^3}\Bigg[2a_{\omega,n} - b_{\omega,n} + \mathcal{L}_{\omega,-}^1\Big(\left(\lambda-1\right)\mathcal{T}_\omega^2 + \left(\lambda+1\right)\mathcal{T}_\omega^3\Big) + \mathcal{L}_{\omega,-}^2\Big(2\left(\mathcal{T}_\omega^3-1\right) - \left(\lambda-1\right)\mathcal{T}_\omega^1 \Big) \notag \\ 
    &\qquad \qquad \qquad \qquad \qquad \qquad \qquad \qquad - \mathcal{L}_{\omega,-}^3\Big(\left(\lambda+1\right)\mathcal{T}_\omega^1 + 2\left(\mathcal{T}_\omega^2-1\right) \Big) \Bigg], \notag \\
    \mathbb{t}_{\omega,d_2} &= \frac{ie^{i\omega\tau}}{3\omega^3}\Bigg[3a_{\omega,n} + \mathcal{L}_{\omega,-}^1\Big(\left(\lambda+1\right)\left(\mathcal{T}_\omega^2 + 2\mathcal{T}_\omega^3\right) -2 \Big) + \mathcal{L}_{\omega,-}^2\Big(\left(\lambda+1\right)\left(\mathcal{T}_\omega^3-\mathcal{T}_\omega^1\right) \Big) \notag \\
    &\qquad \qquad \qquad \qquad \qquad \qquad \qquad \qquad - \mathcal{L}_{\omega,-}^3\Big(\left(\lambda+1\right)\left(\mathcal{T}_\omega^2 + 2\mathcal{T}_\omega^1\right) -2 \Big) \Bigg],
\end{align}
and
\begin{align}
    a_{\omega,n} = i\sqrt{\frac{2}{\pi}}{\lambda \left(1-\lambda ^2\right)^{3/4} \omega ^{3/2}}\mathcal{A}_{\omega,n}\,, \notag \\
    b_{\omega,n} = i\sqrt{\frac{2}{\pi}}{\lambda \left(1-\lambda ^2\right)^{3/4} \omega ^{3/2}}\mathcal{B}_{\omega,n}\,.
\end{align}
Then the transformed amplitudes in the continuous coordinates become
\begin{align} \label{Eq:tamp_asym}
    \widetilde{\mathcal{L}}_{\omega,+}^3 &= \frac{a_{\omega,n}}{3-\lambda} + \frac{(5-\lambda)b_{\omega,n}}{(\lambda-3)(\lambda+1)} + \frac{2(\lambda-1)\widetilde{\mathcal{L}}_{\omega,-}^1}{(\lambda-3)(\lambda+1)} - \frac{2\widetilde{\mathcal{L}}_{\omega,-}^2}{\lambda-3} + \frac{(\lambda-1)^2\widetilde{\mathcal{L}}_{\omega,-}^3}{(\lambda-3)(\lambda+1)}, \notag \\
    \widetilde{\mathcal{L}}_{\omega,+}^2 &= \frac{2a_{\omega,n}}{\lambda-3} + \frac{b_{\omega,n}}{(3-\lambda)} + \frac{2\widetilde{\mathcal{L}}_{\omega,-}^1}{(3-\lambda)} - \frac{(\lambda+1)\widetilde{\mathcal{L}}_{\omega,-}^2}{(3-\lambda)} + \frac{2\widetilde{\mathcal{L}}_{\omega,-}^3}{(3-\lambda)}, \notag \\
    \widetilde{\mathcal{L}}_{\omega,+}^1 &= \frac{a_{\omega,n}}{3-\lambda} + \frac{2(\lambda-2)b_{\omega,n}}{(\lambda-3)(\lambda+1)} + \frac{(\lambda-1)^2\widetilde{\mathcal{L}}_{\omega,-}^1}{(\lambda-3)(\lambda+1)} - \frac{2\widetilde{\mathcal{L}}_{\omega,-}^2}{\lambda-3} + \frac{2(\lambda-1)\widetilde{\mathcal{L}}_{\omega,-}^3}{(\lambda-3)(\lambda+1)},
\end{align}
One can readily check that these transformed amplitudes also satisfy the energy conservation
\begin{equation}
    \widetilde{\mathcal{L}}_{\omega,+}^1 + \widetilde{\mathcal{L}}_{\omega,+}^2 + \widetilde{\mathcal{L}}_{\omega,+}^3 = \widetilde{\mathcal{L}}_{\omega,-}^1 + \widetilde{\mathcal{L}}_{\omega,-}^2 + \widetilde{\mathcal{L}}_{\omega,-}^3.
\end{equation}
Proceeding further, from Eq.~\eqref{Eq:tamp_asym} we can extract the $3 \times 3$ matrix $\mathcal{S}$ corresponding to the asymmetric solution~\eqref{Eq:asym_sol} as follows
\begin{equation} \label{Eq:S_asym}
    \mathcal{S} =\begin{pmatrix}
   \frac{(\lambda-1)^2}{(\lambda-3)(\lambda+1)} & \frac{2}{3-\lambda} & \frac{2(\lambda-1)}{(\lambda-3)(\lambda+1)} \\
\frac{2}{3-\lambda} & \frac{\lambda+1}{\lambda-3} & \frac{2}{3-\lambda} \\
\frac{2(\lambda-1)}{(\lambda-3)(\lambda+1)} & \frac{2}{3-\lambda} & \frac{(\lambda-1)^2}{(\lambda-3)(\lambda+1)}
\end{pmatrix}.
\end{equation}
Note that for $0<\lambda<1$ the $\mathcal{S}$ matrix has eigenvalues with modulus larger than $1$. Hence, this solution is unphysical as it violates the unitarity bounds. The other asymmetric solution, which at $\mathcal{O}(\epsilon^0)$ is given by 
\begin{equation} \label{Eq:asym_sol_other}
    \tau_d =0\,, \quad \sigma_d=0\,, \quad x_d=-4x_s\,, \quad x_s=\frac{\sigma \tilde{p}_{\lambda}}{3}\,,
\end{equation}
where $\tilde{p}_{\lambda}=\frac{\lambda}{\sqrt{1-\lambda^2}}$ and $0 < \lambda < 1$ as before,
also results in the same matrix~\eqref{Eq:S_asym}, and consequently we discard both of these solutions. Similarly, all permutation asymmetric solutions give rise to non-unitary scattering for $n>3$.

{ Nevertheless, non-unitary conformal binary interfaces in unitary CFTs have recently been studied in the literature \cite{Tang:2026ccf}. It would be interesting to see if explicit constructions of non-unitary conformal multi-interfaces can reproduce holographic results.}
    
\end{appendix}

\bibliographystyle{JHEP}
\bibliography{References}

\end{document}